\documentclass[a4paper,12pt]{article}
\usepackage{amsfonts}
\usepackage{amsmath}
\usepackage{amssymb}
\usepackage{graphicx}

\makeatletter
 
\def\diagram{\m@th\leftwidth=\z@ \rightwidth=\z@ \topheight=\z@
\botheight=\z@ \setbox\@picbox\hbox\bgroup}
 
\def\enddiagram{\egroup\wd\@picbox\rightwidth\unitlength
\ht\@picbox\topheight\unitlength \dp\@picbox\botheight\unitlength
\hskip\leftwidth\unitlength\box\@picbox}
 
\def\bfig{\begin{diagram}}
\def\efig{\end{diagram}}
\newcount\wideness \newcount\leftwidth \newcount\rightwidth
\newcount\highness \newcount\topheight \newcount\botheight
 
\def\ratchet#1#2{\ifnum#1<#2 \global #1=#2 \fi}
 
\def\putbox(#1,#2)#3{%
\horsize{\wideness}{#3} \divide\wideness by 2
{\advance\wideness by #1 \ratchet{\rightwidth}{\wideness}}
{\advance\wideness by -#1 \ratchet{\leftwidth}{\wideness}}
\vertsize{\highness}{#3} \divide\highness by 2
{\advance\highness by #2 \ratchet{\topheight}{\highness}}
{\advance\highness by -#2 \ratchet{\botheight}{\highness}}
\put(#1,#2){\makebox(0,0){$#3$}}}
 
\def\putlbox(#1,#2)#3{%
\horsize{\wideness}{#3}
{\advance\wideness by #1 \ratchet{\rightwidth}{\wideness}}
{\ratchet{\leftwidth}{-#1}}
\vertsize{\highness}{#3} \divide\highness by 2
{\advance\highness by #2 \ratchet{\topheight}{\highness}}
{\advance\highness by -#2 \ratchet{\botheight}{\highness}}
\put(#1,#2){\makebox(0,0)[l]{$#3$}}}
 
\def\putrbox(#1,#2)#3{%
\horsize{\wideness}{#3}
{\ratchet{\rightwidth}{#1}}
{\advance\wideness by -#1 \ratchet{\leftwidth}{\wideness}}
\vertsize{\highness}{#3} \divide\highness by 2
{\advance\highness by #2 \ratchet{\topheight}{\highness}}
{\advance\highness by -#2 \ratchet{\botheight}{\highness}}
\put(#1,#2){\makebox(0,0)[r]{$#3$}}}

\def\adjust[#1]{} 
 
\newcount \coefa
\newcount \coefb
\newcount \coefc
\newcount\tempcounta
\newcount\tempcountb
\newcount\tempcountc
\newcount\tempcountd
\newcount\xext
\newcount\yext
\newcount\xoff
\newcount\yoff
\newcount\gap%
\newcount\arrowtypea
\newcount\arrowtypeb
\newcount\arrowtypec
\newcount\arrowtyped
\newcount\arrowtypee
\newcount\height
\newcount\width
\newcount\xpos
\newcount\ypos
\newcount\run
\newcount\rise
\newcount\arrowlength
\newcount\halflength
\newcount\arrowtype
\newdimen\tempdimen
\newdimen\xlen
\newdimen\ylen
\newsavebox{\tempboxa}%
\newsavebox{\tempboxb}%
\newsavebox{\tempboxc}%
 
\newdimen\w@dth
 
\def\setw@dth#1#2{\setbox\z@\hbox{\m@th$#1$}\w@dth=\wd\z@
\setbox\@ne\hbox{\m@th$#2$}\ifnum\w@dth<\wd\@ne \w@dth=\wd\@ne \fi
\advance\w@dth by 1.2em}
 
\def\t@^#1_#2{\allowbreak\def\n@one{#1}\def\n@two{#2}\mathrel
{\setw@dth{#1}{#2}
\mathop{\hbox to \w@dth{\rightarrowfill}}\limits
\ifx\n@one\empty\else ^{\box\z@}\fi
\ifx\n@two\empty\else _{\box\@ne}\fi}}
\def\t@@^#1{\@ifnextchar_{\t@^{#1}}{\t@^{#1}_{}}}
\def\to{\@ifnextchar^{\t@@}{\t@@^{}}}
 
\def\t@left^#1_#2{\def\n@one{#1}\def\n@two{#2}\mathrel{\setw@dth{#1}{#2}
\mathop{\hbox to \w@dth{\leftarrowfill}}\limits
\ifx\n@one\empty\else ^{\box\z@}\fi
\ifx\n@two\empty\else _{\box\@ne}\fi}}
\def\t@@left^#1{\@ifnextchar_{\t@left^{#1}}{\t@left^{#1}_{}}}
\def\toleft{\@ifnextchar^{\t@@left}{\t@@left^{}}}
 
\def\two@^#1_#2{\allowbreak
\def\n@one{#1}\def\n@two{#2}\mathrel{\setw@dth{#1}{#2}
\mathop{\vcenter{\lineskip\z@\baselineskip\z@
                 \hbox to \w@dth{\rightarrowfill}%
                 \hbox to \w@dth{\rightarrowfill}}%
       }\limits
\ifx\n@one\empty\else ^{\box\z@}\fi
\ifx\n@two\empty\else _{\box\@ne}\fi}}
\def\tw@@^#1{\@ifnextchar _{\two@^{#1}}{\two@^{#1}_{}}}
\def\two{\@ifnextchar ^{\tw@@}{\tw@@^{}}}
 
\def\tofr@^#1_#2{\def\n@one{#1}\def\n@two{#2}\mathrel{\setw@dth{#1}{#2}
\mathop{\vcenter{\hbox to \w@dth{\rightarrowfill}\kern-1.7ex
                 \hbox to \w@dth{\leftarrowfill}}%
       }\limits
\ifx\n@one\empty\else ^{\box\z@}\fi
\ifx\n@two\empty\else _{\box\@ne}\fi}}
\def\t@fr@^#1{\@ifnextchar_ {\tofr@^{#1}}{\tofr@^{#1}_{}}}
\def\tofro{\@ifnextchar^ {\t@fr@}{\t@fr@^{}}}

\def\mon{\mathop{\m@th\hbox to
      14.6\P@{\lasyb\char'51\hskip-2.1\P@$\arrext$\hss
$\mathord\rightarrow$}}\limits} 
\def\leftmono{\mathrel{\m@th\hbox to
14.6\P@{$\mathord\leftarrow$\hss$\arrext$\hskip-2.1\P@\lasyb\char'50%
}}\limits} 
\mathchardef\arrext="0200       

\def\settypes(#1,#2,#3){\arrowtypea#1 \arrowtypeb#2 \arrowtypec#3}
\def\settoheight#1#2{\setbox\@tempboxa\hbox{#2}#1\ht\@tempboxa\relax}%
\def\settodepth#1#2{\setbox\@tempboxa\hbox{#2}#1\dp\@tempboxa\relax}%
\def\settokens`#1`#2`#3`#4`{%
     \def\tokena{#1}\def\tokenb{#2}\def\tokenc{#3}\def\tokend{#4}}
\def\setsqparms[#1`#2`#3`#4;#5`#6]{%
\arrowtypea #1
\arrowtypeb #2
\arrowtypec #3
\arrowtyped #4
\width #5
\height #6
}
\def\setpos(#1,#2){\xpos=#1 \ypos#2}

\def\settriparms[#1`#2`#3;#4]{\settripairparms[#1`#2`#3`1`1;#4]}%
 
\def\settripairparms[#1`#2`#3`#4`#5;#6]{%
\arrowtypea #1
\arrowtypeb #2
\arrowtypec #3
\arrowtyped #4
\arrowtypee #5
\width #6
\height #6
}
 
\def\resetparms{\settripairparms[1`1`1`1`1;500]\width 500}
 
\resetparms
 
\def\mvector(#1,#2)#3{
\put(0,0){\vector(#1,#2){#3}}%
\put(0,0){\vector(#1,#2){26}}%
}
\def\evector(#1,#2)#3{{
\arrowlength #3
\put(0,0){\vector(#1,#2){\arrowlength}}%
\advance \arrowlength by-30
\put(0,0){\vector(#1,#2){\arrowlength}}%
}}
 
\def\horsize#1#2{%
\settowidth{\tempdimen}{$#2$}%
#1=\tempdimen
\divide #1 by\unitlength
}
 
\def\vertsize#1#2{%
\settoheight{\tempdimen}{$#2$}%
#1=\tempdimen
\settodepth{\tempdimen}{$#2$}%
\advance #1 by\tempdimen
\divide #1 by\unitlength
}
 
\def\putvector(#1,#2)(#3,#4)#5#6{{%
\ifnum3<\arrowtype
\putdashvector(#1,#2)(#3,#4)#5\arrowtype
\else
\ifnum\arrowtype<-3
\putdashvector(#1,#2)(#3,#4)#5\arrowtype
\else
\xpos=#1
\ypos=#2
\run=#3
\rise=#4
\arrowlength=#5
\ifnum \arrowtype<0
    \ifnum \run=0
        \advance \ypos by-\arrowlength
    \else
        \tempcounta \arrowlength
        \multiply \tempcounta by\rise
        \divide \tempcounta by\run
        \ifnum\run>0
            \advance \xpos by\arrowlength
            \advance \ypos by\tempcounta
        \else
            \advance \xpos by-\arrowlength
            \advance \ypos by-\tempcounta
        \fi
    \fi
    \multiply \arrowtype by-1
    \multiply \rise by-1
    \multiply \run by-1
\fi
\ifcase \arrowtype
\or \put(\xpos,\ypos){\vector(\run,\rise){\arrowlength}}%
\or \put(\xpos,\ypos){\mvector(\run,\rise)\arrowlength}%
\or \put(\xpos,\ypos){\evector(\run,\rise){\arrowlength}}%
\fi\fi\fi
}}
 
\def\putsplitvector(#1,#2)#3#4{
\xpos #1
\ypos #2
\arrowtype #4
\halflength #3
\arrowlength #3
\gap 140
\advance \halflength by-\gap
\divide \halflength by2
\ifnum\arrowtype>0
   \ifcase \arrowtype
   \or \put(\xpos,\ypos){\line(0,-1){\halflength}}%
       \advance\ypos by-\halflength
       \advance\ypos by-\gap
       \put(\xpos,\ypos){\vector(0,-1){\halflength}}%
   \or \put(\xpos,\ypos){\line(0,-1)\halflength}%
       \put(\xpos,\ypos){\vector(0,-1)3}%
       \advance\ypos by-\halflength
       \advance\ypos by-\gap
       \put(\xpos,\ypos){\vector(0,-1){\halflength}}%
   \or \put(\xpos,\ypos){\line(0,-1)\halflength}%
       \advance\ypos by-\halflength
       \advance\ypos by-\gap
       \put(\xpos,\ypos){\evector(0,-1){\halflength}}%
   \fi
\else \arrowtype=-\arrowtype
   \ifcase\arrowtype
   \or \advance \ypos by-\arrowlength
       \put(\xpos,\ypos){\line(0,1){\halflength}}%
       \advance\ypos by\halflength
       \advance\ypos by\gap
       \put(\xpos,\ypos){\vector(0,1){\halflength}}%
   \or \advance \ypos by-\arrowlength
       \put(\xpos,\ypos){\line(0,1)\halflength}%
       \put(\xpos,\ypos){\vector(0,1)3}%
       \advance\ypos by\halflength
       \advance\ypos by\gap
       \put(\xpos,\ypos){\vector(0,1){\halflength}}%
   \or \advance \ypos by-\arrowlength
       \put(\xpos,\ypos){\line(0,1)\halflength}%
       \advance\ypos by\halflength
       \advance\ypos by\gap
       \put(\xpos,\ypos){\evector(0,1){\halflength}}%
   \fi
\fi
}
 
\def\putmorphism(#1)(#2,#3)[#4`#5`#6]#7#8#9{{%
\run #2
\rise #3
\ifnum\rise=0
  \puthmorphism(#1)[#4`#5`#6]{#7}{#8}#9%
\else\ifnum\run=0
  \putvmorphism(#1)[#4`#5`#6]{#7}{#8}#9%
\else
\setpos(#1)%
\arrowlength #7
\arrowtype #8
\ifnum\run=0
\else\ifnum\rise=0
\else
\ifnum\run>0
    \coefa=1
\else
   \coefa=-1
\fi
\ifnum\arrowtype>0
   \coefb=0
   \coefc=-1
\else
   \coefb=\coefa
   \coefc=1
   \arrowtype=-\arrowtype
\fi
\width=2
\multiply \width by\run
\divide \width by\rise
\ifnum \width<0  \width=-\width\fi
\advance\width by60
\if l#9 \width=-\width\fi
\putbox(\xpos,\ypos){#4}
{\multiply \coefa by\arrowlength
\advance\xpos by\coefa
\multiply \coefa by\rise
\divide \coefa by\run
\advance \ypos by\coefa
\putbox(\xpos,\ypos){#5} }%
{\multiply \coefa by\arrowlength
\divide \coefa by2
\advance \xpos by\coefa
\advance \xpos by\width
\multiply \coefa by\rise
\divide \coefa by\run
\advance \ypos by\coefa
\if l#9%
   \putrbox(\xpos,\ypos){#6}%
\else\if r#9%
   \putlbox(\xpos,\ypos){#6}%
\fi\fi }%
{\multiply \rise by-\coefc
\multiply \run by-\coefc
\multiply \coefb by\arrowlength
\advance \xpos by\coefb
\multiply \coefb by\rise
\divide \coefb by\run
\advance \ypos by\coefb
\multiply \coefc by70
\advance \ypos by\coefc
\multiply \coefc by\run
\divide \coefc by\rise
\advance \xpos by\coefc
\multiply \coefa by140
\multiply \coefa by\run
\divide \coefa by\rise
\advance \arrowlength by\coefa
\ifcase\arrowtype
\or \put(\xpos,\ypos){\vector(\run,\rise){\arrowlength}}%
\or \put(\xpos,\ypos){\mvector(\run,\rise){\arrowlength}}%
\or \put(\xpos,\ypos){\evector(\run,\rise){\arrowlength}}%
\fi}\fi\fi\fi\fi}}

\newcount\numbdashes \newcount\lengthdash \newcount\increment
 
\def\howmanydashes{
\numbdashes=\arrowlength \lengthdash=40
\divide\numbdashes by \lengthdash
\lengthdash=\arrowlength
\divide\lengthdash by \numbdashes
\increment=\lengthdash
\multiply\lengthdash by 3
\divide\lengthdash by 5
}
 
\def\putdashvector(#1)(#2,#3)#4#5{%
\ifnum#3=0 \putdashhvector(#1){#4}#5
\else
\ifnum#2=0
\putdashvvector(#1){#4}#5\fi\fi}
 
\def\putdashhvector(#1,#2)#3#4{{%
\arrowlength=#3 \howmanydashes
\multiput(#1,#2)(\increment,0){\numbdashes}%
{\vrule height .4pt width \lengthdash\unitlength}
\arrowtype=#4 \xpos=#1
\ifnum\arrowtype<0 \advance\arrowtype by 7 \fi
\ifcase\arrowtype
\or \advance\xpos by 10
    \put(\xpos,#2){\vector(-1,0){\lengthdash}}
    \advance\xpos by 40
    \put(\xpos,#2){\vector(-1,0){\lengthdash}}
\or \advance \xpos by 10
    \put(\xpos,#2){\vector(-1,0){\lengthdash}}
    \advance\xpos by  \arrowlength
    \advance\xpos by  -50
    \put(\xpos,#2){\vector(-1,0){\lengthdash}}
\or \advance\xpos by 10
    \put(\xpos,#2){\vector(-1,0){\lengthdash}}
\or \advance\xpos by \arrowlength
    \advance\xpos by -\lengthdash
    \put(\xpos,#2){\vector(1,0){\lengthdash}}
\or {\advance\xpos by 10
    \put(\xpos,#2){\vector(1,0){\lengthdash}}}
    \advance\xpos by \arrowlength
    \advance\xpos by -\lengthdash
    \put(\xpos,#2){\vector(1,0){\lengthdash}}
\or \advance\xpos by \arrowlength
    \advance\xpos by -\lengthdash
    \put(\xpos,#2){\vector(1,0){\lengthdash}}
    \advance\xpos by -40
    \put(\xpos,#2){\vector(1,0){\lengthdash}}
   \fi
}}
 
\def\putdashvvector(#1,#2)#3#4{{%
\arrowlength=#3 \howmanydashes
\ypos=#2 \advance\ypos by -\arrowlength
\multiput(#1,#2)(0,\increment){\numbdashes}%
    {\vrule width .4pt height \lengthdash\unitlength}
\arrowtype=#4 \ypos=#2
\ifnum\arrowtype<0 \advance\arrowtype by 7 \fi
\ifcase\arrowtype
\or \advance\ypos by \arrowlength \advance\ypos by -40
    \put(#1,\ypos){\vector(0,1){\lengthdash}}
    \advance\ypos by -40
    \put(#1,\ypos){\vector(0,1){\lengthdash}}
\or \advance\ypos by 10
    \put(#1,\ypos){\vector(0,1){\lengthdash}}
    \advance\ypos by \arrowlength \advance\ypos by -40
    \put(#1,\ypos){\vector(0,1){\lengthdash}}
\or \advance\ypos by \arrowlength \advance\ypos by -40
    \put(#1,\ypos){\vector(0,1){\lengthdash}}
\or \advance\ypos by 10
    \put(#1,\ypos){\vector(0,-1){\lengthdash}}
\or \advance\ypos by 10
    \put(#1,\ypos){\vector(0,-1){\lengthdash}}
    \advance\ypos by \arrowlength \advance\ypos by -40
    \put(#1,\ypos){\vector(0,-1){\lengthdash}}
\or \advance\ypos by 10
    \put(#1,\ypos){\vector(0,-1){\lengthdash}}
    \advance\ypos by 40
    \put(#1,\ypos){\vector(0,-1){\lengthdash}}
\fi
}}
 
\def\puthmorphism(#1,#2)[#3`#4`#5]#6#7#8{{%
\xpos #1
\ypos #2
\width #6
\arrowlength #6
\arrowtype=#7
\putbox(\xpos,\ypos){#3\vphantom{#4}}%
{\advance \xpos by\arrowlength
\putbox(\xpos,\ypos){\vphantom{#3}#4}}%
\horsize{\tempcounta}{#3}%
\horsize{\tempcountb}{#4}%
\divide \tempcounta by2
\divide \tempcountb by2
\advance \tempcounta by30
\advance \tempcountb by30
\advance \xpos by\tempcounta
\advance \arrowlength by-\tempcounta
\advance \arrowlength by-\tempcountb
\putvector(\xpos,\ypos)(1,0)\arrowlength\arrowtype
\divide \arrowlength by2
\advance \xpos by\arrowlength
\vertsize{\tempcounta}{#5}%
\divide\tempcounta by2
\advance \tempcounta by20
\if a#8 %
   \advance \ypos by\tempcounta
   \putbox(\xpos,\ypos){#5}%
\else
   \advance \ypos by-\tempcounta
   \putbox(\xpos,\ypos){#5}%
\fi}}
 
\def\putvmorphism(#1,#2)[#3`#4`#5]#6#7#8{{%
\xpos #1
\ypos #2
\arrowlength #6
\arrowtype #7
\settowidth{\xlen}{$#5$}%
\putbox(\xpos,\ypos){#3}%
{\advance \ypos by-\arrowlength
\putbox(\xpos,\ypos){#4}}%
{\advance\arrowlength by-140
\advance \ypos by-70
\ifdim\xlen>0pt
   \if m#8%
      \putsplitvector(\xpos,\ypos)\arrowlength\arrowtype
   \else
   \putvector(\xpos,\ypos)(0,-1)\arrowlength\arrowtype
   \fi
\else
   \putvector(\xpos,\ypos)(0,-1)\arrowlength\arrowtype
\fi}%
\ifdim\xlen>0pt
   \divide \arrowlength by2
   \advance\ypos by-\arrowlength
   \if l#8%
      \advance \xpos by-40
      \putrbox(\xpos,\ypos){#5}%
   \else\if r#8%
      \advance \xpos by40
      \putlbox(\xpos,\ypos){#5}%
   \else
      \putbox(\xpos,\ypos){#5}%
   \fi\fi
\fi
}}
 
\def\putsquarep<#1>(#2)[#3;#4`#5`#6`#7]{{%
\setsqparms[#1]%
\setpos(#2)%
\settokens`#3`%
\puthmorphism(\xpos,\ypos)[\tokenc`\tokend`{#7}]{\width}{\arrowtyped}b%
\advance\ypos by \height
\puthmorphism(\xpos,\ypos)[\tokena`\tokenb`{#4}]{\width}{\arrowtypea}a%
\putvmorphism(\xpos,\ypos)[``{#5}]{\height}{\arrowtypeb}l%
\advance\xpos by \width
\putvmorphism(\xpos,\ypos)[``{#6}]{\height}{\arrowtypec}r%
}}
 
\def\putsquare{\@ifnextchar <{\putsquarep}{\putsquarep%
   <\arrowtypea`\arrowtypeb`\arrowtypec`\arrowtyped;\width`\height>}}
\def\square{\@ifnextchar< {\squarep}{\squarep
   <\arrowtypea`\arrowtypeb`\arrowtypec`\arrowtyped;\width`\height>}}
\def\squarep<#1>[#2`#3`#4`#5;#6`#7`#8`#9]{{
\setsqparms[#1]
\diagram
\putsquarep<\arrowtypea`\arrowtypeb`\arrowtypec`
\arrowtyped;\width`\height>
(0,0)[#2`#3`#4`{#5};#6`#7`#8`{#9}]
\enddiagram
}}                                                 
\def\putptrianglep<#1>(#2,#3)[#4`#5`#6;#7`#8`#9]{{%
\settriparms[#1]%
\xpos=#2 \ypos=#3
\advance\ypos by \height
\puthmorphism(\xpos,\ypos)[#4`#5`{#7}]{\height}{\arrowtypea}a%
\putvmorphism(\xpos,\ypos)[`#6`{#8}]{\height}{\arrowtypeb}l%
\advance\xpos by\height
\putmorphism(\xpos,\ypos)(-1,-1)[``{#9}]{\height}{\arrowtypec}r%
}}
 
\def\putptriangle{\@ifnextchar <{\putptrianglep}{\putptrianglep
   <\arrowtypea`\arrowtypeb`\arrowtypec;\height>}}
\def\ptriangle{\@ifnextchar <{\ptrianglep}{\ptrianglep
   <\arrowtypea`\arrowtypeb`\arrowtypec;\height>}}
\def\ptrianglep<#1>[#2`#3`#4;#5`#6`#7]{{
\settriparms[#1]
\diagram
\putptrianglep<\arrowtypea`\arrowtypeb`
\arrowtypec;\height>
(0,0)[#2`#3`#4;#5`#6`{#7}]
\enddiagram
}}                                            
 
\def\putqtrianglep<#1>(#2,#3)[#4`#5`#6;#7`#8`#9]{{%
\settriparms[#1]%
\xpos=#2 \ypos=#3
\advance\ypos by\height
\puthmorphism(\xpos,\ypos)[#4`#5`{#7}]{\height}{\arrowtypea}a%
\putmorphism(\xpos,\ypos)(1,-1)[``{#8}]{\height}{\arrowtypeb}l%
\advance\xpos by\height
\putvmorphism(\xpos,\ypos)[`#6`{#9}]{\height}{\arrowtypec}r%
}}
 
\def\putqtriangle{\@ifnextchar <{\putqtrianglep}{\putqtrianglep
   <\arrowtypea`\arrowtypeb`\arrowtypec;\height>}}
\def\qtriangle{\@ifnextchar <{\qtrianglep}{\qtrianglep
   <\arrowtypea`\arrowtypeb`\arrowtypec;\height>}}
\def\qtrianglep<#1>[#2`#3`#4;#5`#6`#7]{{
\settriparms[#1]
\width=\height                                
\diagram
\putqtrianglep<\arrowtypea`\arrowtypeb`
\arrowtypec;\height>
(0,0)[#2`#3`#4;#5`#6`{#7}]
\enddiagram
}}
 
\def\putdtrianglep<#1>(#2,#3)[#4`#5`#6;#7`#8`#9]{{%
\settriparms[#1]%
\xpos=#2 \ypos=#3
\puthmorphism(\xpos,\ypos)[#5`#6`{#9}]{\height}{\arrowtypec}b%
\advance\xpos by \height \advance\ypos by\height
\putmorphism(\xpos,\ypos)(-1,-1)[``{#7}]{\height}{\arrowtypea}l%
\putvmorphism(\xpos,\ypos)[#4``{#8}]{\height}{\arrowtypeb}r%
}}
 
\def\putdtriangle{\@ifnextchar <{\putdtrianglep}{\putdtrianglep
   <\arrowtypea`\arrowtypeb`\arrowtypec;\height>}}
\def\dtriangle{\@ifnextchar <{\dtrianglep}{\dtrianglep
   <\arrowtypea`\arrowtypeb`\arrowtypec;\height>}}
\def\dtrianglep<#1>[#2`#3`#4;#5`#6`#7]{{
\settriparms[#1]
\width=\height                                
\diagram
\putdtrianglep<\arrowtypea`\arrowtypeb`
\arrowtypec;\height>
(0,0)[#2`#3`#4;#5`#6`{#7}]
\enddiagram
}}
 
\def\putbtrianglep<#1>(#2,#3)[#4`#5`#6;#7`#8`#9]{{%
\settriparms[#1]%
\xpos=#2 \ypos=#3
\puthmorphism(\xpos,\ypos)[#5`#6`{#9}]{\height}{\arrowtypec}b%
\advance\ypos by\height
\putmorphism(\xpos,\ypos)(1,-1)[``{#8}]{\height}{\arrowtypeb}r%
\putvmorphism(\xpos,\ypos)[#4``{#7}]{\height}{\arrowtypea}l%
}}
 
\def\putbtriangle{\@ifnextchar <{\putbtrianglep}{\putbtrianglep
   <\arrowtypea`\arrowtypeb`\arrowtypec;\height>}}
\def\btriangle{\@ifnextchar <{\btrianglep}{\btrianglep
   <\arrowtypea`\arrowtypeb`\arrowtypec;\height>}}
\def\btrianglep<#1>[#2`#3`#4;#5`#6`#7]{{
\settriparms[#1]
\width=\height                               
\diagram
\putbtrianglep<\arrowtypea`\arrowtypeb`
\arrowtypec;\height>
(0,0)[#2`#3`#4;#5`#6`{#7}]
\enddiagram
}}
 
\def\putAtrianglep<#1>(#2,#3)[#4`#5`#6;#7`#8`#9]{{%
\settriparms[#1]%
\xpos=#2 \ypos=#3
{\multiply \height by2
\puthmorphism(\xpos,\ypos)[#5`#6`{#9}]{\height}{\arrowtypec}b}%
\advance\xpos by\height \advance\ypos by\height
\putmorphism(\xpos,\ypos)(-1,-1)[#4``{#7}]{\height}{\arrowtypea}l%
\putmorphism(\xpos,\ypos)(1,-1)[``{#8}]{\height}{\arrowtypeb}r%
}}
 
\def\putAtriangle{\@ifnextchar <{\putAtrianglep}{\putAtrianglep
   <\arrowtypea`\arrowtypeb`\arrowtypec;\height>}}
\def\Atriangle{\@ifnextchar <{\Atrianglep}{\Atrianglep
   <\arrowtypea`\arrowtypeb`\arrowtypec;\height>}}
\def\Atrianglep<#1>[#2`#3`#4;#5`#6`#7]{{
\settriparms[#1]
\width=\height                                     
\diagram
\putAtrianglep<\arrowtypea`\arrowtypeb`
\arrowtypec;\height>
(0,0)[#2`#3`#4;#5`#6`{#7}]
\enddiagram
}}
 
\def\putAtrianglepairp<#1>(#2)[#3;#4`#5`#6`#7`#8]{{%
\settripairparms[#1]%
\setpos(#2)%
\settokens`#3`%
\puthmorphism(\xpos,\ypos)[\tokenb`\tokenc`{#7}]{\height}{\arrowtyped}b%
\advance\xpos by\height
\puthmorphism(\xpos,\ypos)[\phantom{\tokenc}`\tokend`{#8}]%
{\height}{\arrowtypee}b%
\advance\ypos by\height
\putmorphism(\xpos,\ypos)(-1,-1)[\tokena``{#4}]{\height}{\arrowtypea}l%
\putvmorphism(\xpos,\ypos)[``{#5}]{\height}{\arrowtypeb}m%
\putmorphism(\xpos,\ypos)(1,-1)[``{#6}]{\height}{\arrowtypec}r%
}}
 
\def\putAtrianglepair{\@ifnextchar <{\putAtrianglepairp}{\putAtrianglepairp%
   <\arrowtypea`\arrowtypeb`\arrowtypec`\arrowtyped`\arrowtypee;\height>}}
\def\Atrianglepair{\@ifnextchar <{\Atrianglepairp}{\Atrianglepairp%
   <\arrowtypea`\arrowtypeb`\arrowtypec`\arrowtyped`\arrowtypee;\height>}}
 
\def\Atrianglepairp<#1>[#2;#3`#4`#5`#6`#7]{{
\settripairparms[#1]
\settokens`#2`
\width=\height                                
\diagram
\putAtrianglepairp                            
<\arrowtypea`\arrowtypeb`\arrowtypec`
\arrowtyped`\arrowtypee;\height>
(0,0)[{#2};#3`#4`#5`#6`{#7}]
\enddiagram
}}
 
\def\putVtrianglep<#1>(#2,#3)[#4`#5`#6;#7`#8`#9]{{%
\settriparms[#1]%
\xpos=#2 \ypos=#3
\advance\ypos by\height
{\multiply\height by2
\puthmorphism(\xpos,\ypos)[#4`#5`{#7}]{\height}{\arrowtypea}a}%
\putmorphism(\xpos,\ypos)(1,-1)[`#6`{#8}]{\height}{\arrowtypeb}l%
\advance\xpos by\height
\advance\xpos by\height
\putmorphism(\xpos,\ypos)(-1,-1)[``{#9}]{\height}{\arrowtypec}r%
}}
 
\def\putVtriangle{\@ifnextchar <{\putVtrianglep}{\putVtrianglep
   <\arrowtypea`\arrowtypeb`\arrowtypec;\height>}}
\def\Vtriangle{\@ifnextchar <{\Vtrianglep}{\Vtrianglep
   <\arrowtypea`\arrowtypeb`\arrowtypec;\height>}}
\def\Vtrianglep<#1>[#2`#3`#4;#5`#6`#7]{{
\settriparms[#1]
\width=\height                                 
\diagram
\putVtrianglep<\arrowtypea`\arrowtypeb`
\arrowtypec;\height>
(0,0)[#2`#3`#4;#5`#6`{#7}]
\enddiagram
}}
 
\def\putVtrianglepairp<#1>(#2)[#3;#4`#5`#6`#7`#8]{{
\settripairparms[#1]%
\setpos(#2)%
\settokens`#3`%
\advance\ypos by\height
\putmorphism(\xpos,\ypos)(1,-1)[`\tokend`{#6}]{\height}{\arrowtypec}l%
\puthmorphism(\xpos,\ypos)[\tokena`\tokenb`{#4}]{\height}{\arrowtypea}a%
\advance\xpos by\height
\puthmorphism(\xpos,\ypos)[\phantom{\tokenb}`\tokenc`{#5}]%
{\height}{\arrowtypeb}a%
\putvmorphism(\xpos,\ypos)[``{#7}]{\height}{\arrowtyped}m%
\advance\xpos by\height
\putmorphism(\xpos,\ypos)(-1,-1)[``{#8}]{\height}{\arrowtypee}r%
}}
 
\def\putVtrianglepair{\@ifnextchar <{\putVtrianglepairp}{\putVtrianglepairp%
    <\arrowtypea`\arrowtypeb`\arrowtypec`\arrowtyped`\arrowtypee;\height>}}
\def\Vtrianglepair{\@ifnextchar <{\Vtrianglepairp}{\Vtrianglepairp%
    <\arrowtypea`\arrowtypeb`\arrowtypec`\arrowtyped`\arrowtypee;\height>}}
\def\Vtrianglepairp<#1>[#2;#3`#4`#5`#6`#7]{{
\settripairparms[#1]
\settokens`#2`
\diagram
\putVtrianglepairp                             
<\arrowtypea`\arrowtypeb`\arrowtypec`
\arrowtyped`\arrowtypee;\height>
(0,0)[{#2};#3`#4`#5`#6`{#7}]
\enddiagram
}}

\def\putCtrianglep<#1>(#2,#3)[#4`#5`#6;#7`#8`#9]{{%
\settriparms[#1]%
\xpos=#2 \ypos=#3
\advance\ypos by\height
\putmorphism(\xpos,\ypos)(1,-1)[``{#9}]{\height}{\arrowtypec}l%
\advance\xpos by\height
\advance\ypos by\height
\putmorphism(\xpos,\ypos)(-1,-1)[#4`#5`{#7}]{\height}{\arrowtypea}l%
{\multiply\height by 2
\putvmorphism(\xpos,\ypos)[`#6`{#8}]{\height}{\arrowtypeb}r}%
}}
 
\def\putCtriangle{\@ifnextchar <{\putCtrianglep}{\putCtrianglep
    <\arrowtypea`\arrowtypeb`\arrowtypec;\height>}}
\def\Ctriangle{\@ifnextchar <{\Ctrianglep}{\Ctrianglep
    <\arrowtypea`\arrowtypeb`\arrowtypec;\height>}}
\def\Ctrianglep<#1>[#2`#3`#4;#5`#6`#7]{{
\settriparms[#1]
\width=\height                               
\diagram
\putCtrianglep<\arrowtypea`\arrowtypeb`
\arrowtypec;\height>
(0,0)[#2`#3`#4;#5`#6`{#7}]
\enddiagram
}}                                           
\def\putDtrianglep<#1>(#2,#3)[#4`#5`#6;#7`#8`#9]{{%
\settriparms[#1]%
\xpos=#2 \ypos=#3
\advance\xpos by\height \advance\ypos by\height
\putmorphism(\xpos,\ypos)(-1,-1)[``{#9}]{\height}{\arrowtypec}r%
\advance\xpos by-\height \advance\ypos by\height
\putmorphism(\xpos,\ypos)(1,-1)[`#5`{#8}]{\height}{\arrowtypeb}r%
{\multiply\height by 2
\putvmorphism(\xpos,\ypos)[#4`#6`{#7}]{\height}{\arrowtypea}l}%
}}
 
\def\putDtriangle{\@ifnextchar <{\putDtrianglep}{\putDtrianglep
    <\arrowtypea`\arrowtypeb`\arrowtypec;\height>}}
\def\Dtriangle{\@ifnextchar <{\Dtrianglep}{\Dtrianglep
   <\arrowtypea`\arrowtypeb`\arrowtypec;\height>}}
\def\Dtrianglep<#1>[#2`#3`#4;#5`#6`#7]{{
\settriparms[#1]
\width=\height                              
\diagram
\putDtrianglep<\arrowtypea`\arrowtypeb`
\arrowtypec;\height>
(0,0)[#2`#3`#4;#5`#6`{#7}]
\enddiagram
}}                                          
\def\setrecparms[#1`#2]{\width=#1 \height=#2}%
 
\def\recursep<#1`#2>[#3;#4`#5`#6`#7`#8]{{\m@th
\width=#1 \height=#2
\settokens`#3`
\settowidth{\tempdimen}{$\tokena$}
\ifdim\tempdimen=0pt
  \savebox{\tempboxa}{\hbox{$\tokenb$}}%
  \savebox{\tempboxb}{\hbox{$\tokend$}}%
  \savebox{\tempboxc}{\hbox{$#6$}}%
\else
  \savebox{\tempboxa}{\hbox{$\hbox{$\tokena$}\times\hbox{$\tokenb$}$}}%
  \savebox{\tempboxb}{\hbox{$\hbox{$\tokena$}\times\hbox{$\tokend$}$}}%
  \savebox{\tempboxc}{\hbox{$\hbox{$\tokena$}\times\hbox{$#6$}$}}%
\fi
\ypos=\height
\divide\ypos by 2
\xpos=\ypos
\advance\xpos by \width
\bfig
\putCtrianglep<-1`1`1;\ypos>(0,0)[`\tokenc`;#5`#6`{#7}]%
\puthmorphism(\ypos,0)[\tokend`\usebox{\tempboxb}`{#8}]{\width}{-1}b%
\puthmorphism(\ypos,\height)[\tokenb`\usebox{\tempboxa}`{#4}]{\width}{-1}a%
\advance\ypos by \width
\putvmorphism(\ypos,\height)[``\usebox{\tempboxc}]{\height}1r%
\efig
}}
 
\def\recurse{\@ifnextchar <{\recursep}{\recursep<\width`\height>}}
 
\def\puttwohmorphisms(#1,#2)[#3`#4;#5`#6]#7#8#9{{%
\puthmorphism(#1,#2)[#3`#4`]{#7}0a
\ypos=#2
\advance\ypos by 20
\puthmorphism(#1,\ypos)[\phantom{#3}`\phantom{#4}`#5]{#7}{#8}a
\advance\ypos by -40
\puthmorphism(#1,\ypos)[\phantom{#3}`\phantom{#4}`#6]{#7}{#9}b
}}
 
\def\puttwovmorphisms(#1,#2)[#3`#4;#5`#6]#7#8#9{{%
\putvmorphism(#1,#2)[#3`#4`]{#7}0a
\xpos=#1
\advance\xpos by -20
\putvmorphism(\xpos,#2)[\phantom{#3}`\phantom{#4}`#5]{#7}{#8}l
\advance\xpos by 40
\putvmorphism(\xpos,#2)[\phantom{#3}`\phantom{#4}`#6]{#7}{#9}r
}}
 
\def\puthcoequalizer(#1)[#2`#3`#4;#5`#6`#7]#8#9{{%
\setpos(#1)%
\puttwohmorphisms(\xpos,\ypos)[#2`#3;#5`#6]{#8}11%
\advance\xpos by #8
\puthmorphism(\xpos,\ypos)[\phantom{#3}`#4`#7]{#8}1{#9}
}}
 
\def\putvcoequalizer(#1)[#2`#3`#4;#5`#6`#7]#8#9{{%
\setpos(#1)%
\puttwovmorphisms(\xpos,\ypos)[#2`#3;#5`#6]{#8}11%
\advance\ypos by -#8
\putvmorphism(\xpos,\ypos)[\phantom{#3}`#4`#7]{#8}1{#9}
}}
 
\def\putthreehmorphisms(#1)[#2`#3;#4`#5`#6]#7(#8)#9{{%
\setpos(#1) \settypes(#8)
\if a#9 %
     \vertsize{\tempcounta}{#5}%
     \vertsize{\tempcountb}{#6}%
     \ifnum \tempcounta<\tempcountb \tempcounta=\tempcountb \fi
\else
     \vertsize{\tempcounta}{#4}%
     \vertsize{\tempcountb}{#5}%
     \ifnum \tempcounta<\tempcountb \tempcounta=\tempcountb \fi
\fi
\advance \tempcounta by 60
\puthmorphism(\xpos,\ypos)[#2`#3`#5]{#7}{\arrowtypeb}{#9}
\advance\ypos by \tempcounta
\puthmorphism(\xpos,\ypos)[\phantom{#2}`\phantom{#3}`#4]{#7}{\arrowtypea}{#9}
\advance\ypos by -\tempcounta \advance\ypos by -\tempcounta
\puthmorphism(\xpos,\ypos)[\phantom{#2}`\phantom{#3}`#6]{#7}{\arrowtypec}{#9}
}}
 
\def\setarrowtoks[#1`#2`#3`#4`#5`#6]{%
\def\toka{#1}
\def\tokb{#2}
\def\tokc{#3}
\def\tokd{#4}
\def\toke{#5}
\def\tokf{#6}
}
\def\hex{\@ifnextchar <{\hexp}{\hexp<1000`400>}}
\def\hexp<#1`#2>[#3`#4`#5`#6`#7`#8;#9]{%
\setarrowtoks[#9]
\yext=#2 \advance \yext by #2
\xext=#1 \advance\xext by \yext
\bfig
\putCtriangle<-1`0`1;#2>(0,0)[`#5`;\tokb``\tokd]
\xext=#1 \yext=#2 \advance \yext by #2
\putsquare<1`0`0`1;\xext`\yext>(#2,0)[#3`#4`#7`#8;\toka```\tokf]
\advance \xext by #2
\putDtriangle<0`1`-1;#2>(\xext,0)[`#6`;`\tokc`\toke]
\efig
}
\makeatother

\let\ssection=\section
\renewcommand{\section}{\setcounter{equation}{0}\ssection}

\newtheorem{definition}{Definition}[section]
\newtheorem{theorem}{Theorem}[section]
\newtheorem{lemma}[theorem]{Lemma}

\newenvironment{proof}[1][Proof]{\noindent\textbf{#1.} }{\ \rule{0.5em}{0.5em}}

\newcommand\mapright[1]{\smash{
        \mathop{\mbox{\large{$\longrightarrow$}}}\limits^{#1}}}

\newcommand\mathC{\mkern1mu\raise2.2pt\hbox{$\scriptscriptstyle|$}
        {\mkern-7mu\rm C}}            
\newcommand{\mathR}{{\rm I\! R}}      

\newcommand\ie{{i.e.},}

\newcommand{\ga}{\gamma}
\newcommand{\Ga}{\Gamma}

\renewcommand\l{\lambda}
\newcommand\s{\sigma}
\newcommand\Si{\Sigma}
\newcommand\De{\Delta}

\renewcommand{\O}{\Omega}
\newcommand{\id}{{\rm id}}
\newcommand\la{\langle}
\newcommand{\map}{\rightarrow}               
\newcommand\ra{\rangle}

\newcommand{\op}{{\rm op}}           
\newcommand{\picl}{\pi_{{\rm cl}}}   
\newcommand{\piqt}{\pi_{{\rm qt}}}   

\newcommand{\A}{{\hat A}}

\renewcommand{\P}{{\hat P}}
\renewcommand{\S}{{\cal S}}
\newcommand{\Hi}{{\cal H}}

\newcommand\BH{\mathcal{B(H)}}
\newcommand\PH{\mathcal{P(H)}}
\newcommand\PV{\mathcal{P}(V)}

\newcommand\TO{\mathbb{T}}   

\newcommand\LeftDB{[\mkern-3mu[}
\newcommand\RightDB{]\mkern-3mu]}

\newcommand\bra[1]{\langle #1|\,}
\newcommand\ket[1]{\,|#1\rangle}
\newcommand\eq[1]{(\ref{#1})}
\newcommand\eqs[2]{(\ref{#1}--\ref{#2})}
\newcommand\SAin[1]{\mbox{``}A\,\varepsilon\,#1\mbox{''}}
\newcommand\Ain[1]{A\,\varepsilon\,#1}
\newcommand\va[1]{\tilde{#1}}

\newcommand\das[1]{\delta(\hat{#1})}            
\newcommand\dasto[2]{\delta(\hat{#2})_{#1}}     
\newcommand\dastoo[2]{\delta^o(\hat{#2})_{#1}}  
\newcommand\dastoi[2]{\delta^i(\hat{#2})_{#1}}  
\newcommand\dasmap{\delta}                      

\newcommand\dasBo[1]{\breve{\delta}^o(#1)}

\newcommand\q[1]{`#1\mbox{'}}
\renewcommand\L[1]{\mathcal{L}({#1})}
\newcommand\PL[1]{{\cal PL}(#1)}
\renewcommand\sp[1]{{\rm sp}(\hat A)}
\newcommand\GT[1]{\overline {#1}}

\newcommand\Val[1]{\LeftDB\,#1\,\RightDB}
\newcommand\TVal[2]{\nu\big(#1;#2\big)}         
\newcommand\TValM[1]{\nu(\,#1\,)}               

\newcommand\so[1]{#1}                           
\newcommand\name[1]{\ulcorner #1\urcorner}                
\newcommand\cha[1]{\chi_{#1}}                   

\newcommand\ps[1]{\underline{#1}}        
\newcommand{\Om}{\ps{\Omega}}            
\newcommand{\G}{\ps{O}}                   
\renewcommand{\H}{\ps{I}}                 
\newcommand{\dG}{\ps{\mkern1mu\raise2.5pt\hbox{$\scriptscriptstyle|$}
        {\mkern-7mu\rm O}}}                 
\newcommand{\dOU}{\ps{\mkern1mu\raise2.5pt\hbox{$\scriptscriptstyle|$}
        {\mkern-7mu\rm U}}}               

\newcommand{\Sig}{\ps{\Sigma}}            
\newcommand{\PSig}{P_{{\rm cl}}\Sig}      
\newcommand{\R}{{\cal R}}                 

\newcommand{\SR}{\ps{{\mathR}^\succeq}}         

\newcommand\F[1]{F_{\L{#1}}\big(\Sigma,\R\big)}
\newcommand\Ob[1]{{\rm Ob(#1)}}
\newcommand\Sub[1]{{\rm Sub}(#1)}              
\newcommand\Subcl[1]{{\rm Sub}_{{\rm cl}}(#1)} 

\newcommand\Set{{\bf Sets}}                    
\newcommand\SetH[1]{\Set^{{\V{#1}}^{\rm op}}}  
\newcommand\SetC[1]{\Set^{{#1}^{\rm op}}}      

\newcommand\V[1]{{\cal V}(\Hi_{#1})}           

\begin{document}

\begin{titlepage}

\begin{center}
{\large\bf A Topos Foundation for Theories of Physics:

II. Daseinisation and the Liberation of Quantum Theory}
\end{center}

\vspace{0.8 truecm}
\begin{center}
        A.~D\"oring\footnote{email: a.doering@imperial.ac.uk}\\[10pt]

\begin{center}                      and
\end{center}

        C.J.~Isham\footnote{email: c.isham@imperial.ac.uk}\\[10pt]

        The Blackett Laboratory\\ Imperial College of Science,
        Technology \& Medicine\\ South Kensington\\ London SW7 2BZ\\
\end{center}

\begin{center}
       6 March, 2007
\end{center}

\vspace{0.8 truecm}

\begin{abstract}

This paper is the second in a series whose goal is to develop a
fundamentally new way of constructing theories of physics. The
motivation comes from a desire to address certain deep issues that
arise when contemplating quantum theories of space and time.

Our basic contention is that constructing a theory of physics is
equivalent to finding a representation in a topos of a certain
formal language that is attached to the system. Classical physics
arises when the topos is the category of sets. Other types of
theory employ a different topos.

In this paper, we study in depth the topos representation of the
propositional language, $\PL{S}$, for the case of quantum theory.
In doing so, we make a direct link with, and clarify, the earlier
work on applying topos theory to quantum physics. The key step is
a process we term `daseinisation' by which a projection operator
is mapped to a sub-object of the spectral presheaf---the topos
quantum analogue of a classical state space.

In the second part of the paper we change gear with the
introduction of the more sophisticated local language $\L{S}$.
From this point forward, throughout the rest of the series of
papers, our attention will be devoted almost entirely to this
language. In the present paper, we use $\L{S}$ to study `truth
objects' in the topos. These are objects in the topos that play
the role of states: a necessary development as the spectral
presheaf has no global elements, and hence there are no
microstates in the sense of classical physics. Truth objects
therefore play a crucial role in our formalism.
\end{abstract}
\end{titlepage}

\section{Introduction}
This is the second in a series of papers whose aim is to formulate
a general framework for expressing theories of physics in a topos
other than that of sets. In paper I we introduced the idea that a
formal language can be attached to each  system, $S$, and that, in
the broadest sense, constructing a physical theory of $S$ is
equivalent to finding a representation, $\phi$,  of this language
in a topos $\tau_\phi$ \cite{DI(1)}.

It is expected that, for a given system, different theory-types
(such as classical physics, quantum physics, and others yet to be
discovered) will be represented in different topoi. Typically,
more than one system will share the same language, and, for a
given theory-type, these systems will generally be represented in
the same topos. However, the details of the representation will be
system dependent.

For example, let the system, $S$, be a non-relativistic point
particle moving in three dimensions with the Hamiltonian
$H=\underline{p}\cdot \underline{p}/{2m^2}+ V(\underline x)$. The
application of the theory-type ``classical physics'' involves a
representation in the topos, $\Set$, of sets, and different
representations of the language of $S$ correspond to different
choices of the potential $V(\underline{x})$. On the other hand, a
representation of the same system, but for the theory-type
``quantum physics'', employs a topos of presheaves, $\SetC{C}$,
over a certain category $\cal C$ (see below) that depends on the
system. Once again, different Hamiltonians correspond to different
representations in this topos.

In the first paper, I, we discussed two different types of
language that can be attached to a system $S$. The first is a
simple propositional language, $\PL{S}$, that is generated  by
primitive propositions of the form $\SAin\De$. Such a proposition
is to be understood in a realist fashion as asserting  ``The
physical quantity $A$ has a value that lies in the (Borel) subset
$\De$ of $\mathR$''. The language $\PL{S}$ has the logical
connectives $\land,\lor,\neg, \Rightarrow$; but nothing else. In
particular, it does not include the quantifiers
 $\forall$ and $\exists$.
Nevertheless, it does enable compound propositions about the world
to be asserted. It also has a deductive structure that follows
from a set of axioms that ensure the validity of intuitionistic
logic. The choice of intuitionistic logic over against Boolean
logic is made with the hindsight of knowing that  each topos has
an internal logic of this type.

This propositional language is used in the first half of the
present paper where we  explore in detail the representation of
propositions by (clopen) sub-objects of the state object (the
significance of the word `clopen' is explained below).  However,
for the sake of overall clarity it should be emphasised that the
work involving $\PL{S}$ is a something of a side-line to our main
programme, which is concerned with the local language $\L{S}$. In
fact, logically speaking, the reader could jump straight to
Section \ref{Sec:TruthValues}, which deals with the idea of a
`truth object' in the context of the language $\L{S}$. However, we
decided to include the $\PL{S}$-material  because it is what links
most closely to the original work on using topos ideas in quantum
theory. Indeed, as we shall see, the new material on the
representation of $\PL{S}$ includes a vital concept that was not
understood before, and which places the earlier work in a much
clearer light.

The second language  introduced in  paper I is far more powerful.
Firstly, it is higher order, so that the existential connectives
$\forall, \exists$ are included. Secondly, this, so-called,
`local' language, $\L{S}$, is \emph{typed}: a feature that allows
the most important ingredients of any theory of physics to be
included in a very specific way.

Specifically, $\L{S}$ contains the `ground-type symbols', $\Si$
and $\R$, which are construed as the `linguistic precursors' of
the state object, and quantity-value object, respectively. Thus,
in any representation\footnote{A more comprehensive notation is
$\tau_\phi(S)$, which draws attention to the system $S$ under
discussion; similarly, the state object could be written as
$\Sigma_{\phi,S}$, and so on. This extended notation is used in
paper IV where we are concerned with the relations between
\emph{different} systems, and then it is essential to indicate
which system is meant. However, in the present paper, only one
system at a time is being considered, and so the truncated
notation is adequate.}, $\phi$, of $\L{S}$ in a topos $\tau_\phi$,
the symbols $\Si$ and $\R$  are mapped to objects $\Si_\phi$ and
$\R_\phi$ in $\tau_\phi$ which \emph{are} the state object and
quantity-value object, respectively, for this particular theory.

The application of this structure to classical physics was
discussed in I, and  in the present paper, we want to turn to the,
more challenging, case of quantum theory. We know from the earlier
work on applying topos ideas to quantum theory, that  the topos of
the quantum representation, $\phi$, of  $\L{S}$ is the category of
presheaves, $\SetH{}$, over the category $\V{}$ of unital, abelian
von Neumann subalgebras of $\BH$. Here,  $\BH$ is the
non-commutative algebra of all bounded operators on the Hilbert
space, $\Hi$, of the quantum system. In this representation, the
state object, $\Si_\phi$,  is the spectral presheaf\footnote{As a
pedagogical aid, all presheaves will be denoted with an underlined
symbol, like $\Sig$, $\ps\R$, $\Om$ etc.} $\Sig$ that was
discussed at length in the earlier work \cite{IB98, IB99,
IB00,IB02}; see Definition \ref{Def_SpectralPresheaf} in this
paper. This is the quantum analogue of the classical state space.

Still, it remains true that only a limited range of questions can
be addressed using the propositional language, $\PL{S}$, and to
appreciate properly the full  scope of the `toposification' ideas
it is essential to employ the more sophisticated local language
$\L{S}$. For this reason,  in the second part of the paper we
switch to using $\L{S}$, and this will be the focus of our
attention for the remaining papers in this series.

Amongst other things, using $\L{S}$  involves identifying the
quantity-value presheaf: something that is not part of the simpler
language $\PL{S}$. The quantity-value presheaf, $\R_\phi$, and
related topics are discussed in papers III and IV
\cite{DI(3),DI(4)}. In particular, in paper III we show how a
function symbol $A:\Si\map\R$ (\ie\ a physical quantity) can be
represented by an arrow $A_\phi:\Si_\phi\map\R_\phi$, where
$\Si_\phi$ is the spectral presheaf $\Sig$.

Returning to $\PL{S}$, the plan of the present paper is as
follows.  A key step in constructing a topos representation is an
operation that we call `daseinisation', and this is discussed in
detail in Section \ref{Sec:QuPropSpec}. This involves constructing
a map from the lattice of projectors $\PH$ to the Heyting algebra
$\Subcl{\Sig}$ of clopen sub-objects of the spectral presheaf
$\Sig$. The motivation for this construction lies in the earlier
work on topos theory and quantum mechanics
\cite{IB98,IB99,IB00,IB02} but the key technical ingredient is due
to de Groote \cite{deG05}.

One feature of this construction is that there are more clopen
sub-objects of $\Sig$ than those given by daseinisation of
projection operators. This raises the question of whether there
are special, definitive features of the sub-objects that
\emph{are} of the form $\das{P}$, $\P\in\PH$, and this is
discussed in detail in Section \ref{Sec:SpecNatureDasP} where we
show that the daseinised propositions are `optimal' in the lattice
of all sub-objects of $\Sig$.

The state object, $\Sig$, in the quantum topos has no global
elements (this statement is equivalent to the Kochen-Specker
theorem) and, therefore,  the assignation of `truth values' to
propositions needs an approach that is structurally different from
the one used in classical physics. This issue is discussed in
detail in Section \ref{Sec:TruthValues}, using a method that rests
heavily on  the local language $\L{S}$. In particular, we
introduce into $\L{S}$ a variable that is the linguistic precursor
of the truth object needed in the topos representations. We then
use this to motivate an analogous construction in quantum theory
for the propositional language $\PL{S}$. The same kind of truth
object is used in the context of a representation of the local
language $\L{S}$ in quantum theory, see paper III.

\paragraph{A cautionary caveat.} At this point, before embarking
on the main text, we want to remark on the scope of the work
reported in this paper, and in paper III: \ie\  the application of
topos ideas to standard quantum theory via the representation of a
system language.

For the language $\PL{S}$, a key result from the topos
constructions is that, given any quantum state $\ket\psi$, there
are generalised truth values, $\TVal{\Ain\De}{\ket\psi}$, for
propositions $\SAin\De$. These truth values belong to the Heyting
algebra, $\Ga\O_\phi$, of global elements of the sub-object
classifier, $\O_\phi$, of the topos concerned. In making these
assignments, nothing is said about `measurements', or `observers',
or even `probability': there is just the truth value
$\TVal{\Ain\De}{\ket\psi}\in\Ga\O_\phi$.

It is this absence of  instrumentalist concepts that
motivates/justifies the appellation `neo-realist' for this topos
approach to quantum theory.  However, we want to insert a
cautionary remark.  Our specific topos constructions for  quantum
theory are based on the usual mathematical formalism of
self-adjoint operators on a Hilbert space; and some physicists,
such as Bohr,  have asserted that this formalism is
\emph{fundamentally} associated with an instrumentalist
interpretation of the theory, whose predictions are probabilities
of the results of making measurements. If this is true, our
neo-realism would sit uncomfortably on the formalism.

On the other hand, maybe our quantum results \emph{are} meaningful
as they stand---particularly the ones obtained using the local
language $\L{S}$ (in paper III). If so, an  intriguing challenge
would be to use these methods to construct a neo-realist
perspective on a quantum-cosmological model. A particularly
appropriate example is the recent construction by Kessari of  a
(HPO\footnote{There are various mathematical approaches to
formulating consistent-histories quantum theory, but the one to
which our topos work could most easily be adapted is the `History
Projection Operator' (HPO) scheme in which history propositions
are represented by projection operators.}) consistent-histories
version of the Robertson-Walker universe \cite{Kess07}.

Evidently, the operation we have called `daseinisation' is a
bridge between the intrinsic instrumentalism of the standard
quantum formalism,  and a full neo-realism associated with the
language $\L{S}$ which makes no reference  to entities outside
that language. In this sense, standard quantum theory is a
`hybrid' object that interpolates between the instrumentalist
world, with its various background structures, and the neo-realist
world, in which almost everything has a linguistic precursor.

However, as we keep emphasising, our main goal is to construct
tools for developing new types of theory, not just to look at
standard quantum theory from a novel angle. For us, the most
important thing about the topos approach to standard quantum
theory is that (we hope) it serves as a  paradigmatic example of
the types of mathematical structure that might be needed to
develop theories that  go `beyond' standard quantum theory, and
that are not slaves to the use of continuum entities. We will have
more to say about this central theme at various points in this
series of papers.

\section{Quantum Propositions as Sub-Objects of the Spectral Presheaf}
\label{Sec:QuPropSpec}
\subsection{From Projections to Global Sections of the Outer
Presheaf}\label{_SubS_FromProjsToGlobSecs}
\subsubsection{The Definition of $\dasto{V}{P}$}\label{SubSub:defdas}
The fundamental thesis of our work is that, in constructing
theories of physics, one should seek representations of a formal
language in a topos that may be other than $\Set$. We  want now to
study this idea closely in the context of the toposification of
standard quantum theory, with particular emphasis on a topos
representation of propositions. Most `standard' quantum systems
(for example, one-dimensional motion with a Hamiltonian
$H=\frac{p^2}{2m}+V(x)$) are obtained by `quantising' a classical
system, and consequently the language is the same as it is for the
classical system.

As explained in the Introduction, in the first half of this paper
we concentrate   on the propositional language $\PL{S}$, so that
the critical task is to find the map $\piqt:\PL{S}_0\map
\Subcl{\Sig}$, where the primitive propositions in $\PL{S}_0$ are
of the form $\SAin\De$ (the notation is explained in paper I
\cite{DI(1)}). As we  shall see, this is where the critical
concept of \emph{daseinisation} arises: the procedure whereby a
projector $\P$ is transformed to a (clopen) sub-object, $\das{P}$,
of the spectral presheaf in the topos $\SetH{}$.

In standard quantum theory, a physical quantity is represented by
a self-adjoint operator $\hat{A}$ in the algebra, $\BH,$ of all
bounded operators on $\Hi$. If $\De\subseteq\mathR$ is a Borel
subset, we know from the spectral theorem  that the proposition
$\SAin{\De}$ is represented by\footnote{Note, however, that the
map from propositions to projections is not injective: two
propositions $\SAin{\De_1}$ and ``$B\varepsilon\De_{2}$'' about
two distinct physical quantities, $A$ and $B$, can be represented
by the same projector: \ie\ $\hat E[A\in\De_1]=\hat
E[B\in\De_2]$.} a projection operator $\hat E[A\in\De]$ in $\BH$.
For typographical simplicity, for the rest of this Section, $\hat
E[A\in\De]$ will be denoted by $\P$.

We are going to consider the projection operator  $\P$ from the
perspective of the `category of contexts' that is the basis of the
topos approach to quantum theory. There are several possible
choices for this category, and these are considered in detail in
the original papers \cite{IB98,IB99,IB00,IB02}. Mathematically,
they are ultimately all equivalent, but here we have chosen to use
the category $\V{}$ of unital, abelian subalgebras of $\BH$. The
category structure is that of a partially-ordered set whose
objects are the abelian subalgebras, and in which there is an
arrow $i_{V^\prime V}:V^\prime\map V$,
$V^\prime,V\in\Ob{\V{}}$,\footnote{We denote by $\Ob{{\cal C}}$
the collection of all objects in the category $\cal C$.} if and
only if $V^\prime\subseteq V$.  By definition, the trivial
subalgebra $V_0=\mathC\hat{1}$ is not included in the objects of
$\V{}$. A context could also be called a `world-view', or a
`window on the world', or even a
Weltanschauung\footnote{`Weltanschauung' is a splendid German
word. `Welt' means world; `schauen' is a verb and means to look,
to view; `anschauen' is to look at; and `-ung' at the end of a
word can make a noun from a verb. So it's Welt-an-schau-ung.};
mathematicians call it a `stage of truth'.

The critical question is what can be said about the projector $\P$
`from the view-point' of a particular context $V\in\Ob{\V{}}$? If
$\P$ belongs to $V$ then  a `full' image  of $\P$ is obtained from
this view-point, and there is nothing more to say. However,
suppose the abelian subalgebra $V$ does \emph{not} contain $\P$:
what then?

We need to `approximate' $\P$ from the perspective of $V$, and an
important ingredient in our work is to define this as meaning the
`smallest' projection operator, $\dasto{V}{P}$, in $V$ that is
greater than, or equal to, $\P$:
\begin{equation}
\dasto{V}{P}:=\bigwedge\big\{\hat{Q}\in\PV\mid \hat{Q}\succeq
\P\big\}.           \label{Def:dasouter}
\end{equation}
where `$\succeq$' is the usual `quantum-logic' ordering of
projection operators, and where $\PV$ denotes the set of all
projection operators in $V$.

To see what this means, let $\P$ and $\hat Q$ represent the
propositions $\SAin\De$ and $\SAin{\De^\prime}$ respectively with
$\De\subseteq\De^{\prime}$, so that $\P\preceq\hat Q$.
 Since we learn less about the value of
$A$ from the proposition $\SAin{\De^{\prime}}$ than from
$\SAin{\De}$, the former proposition is said to be \emph{weaker}.
Clearly, the weaker proposition $\SAin{\De^{\prime}}$ is implied
by the stronger proposition $\SAin{\De}$.  The construction of
$\dasto{V}{P}$ as the smallest projection in $V$ greater than or
equal to $\P$ thus gives the strongest proposition expressible in
$V$ that is implied by $\P$ (although, if $\hat A\notin V$, the
projection $\dasto{V}{P}$ cannot be interpreted as a proposition
about $A$ in general).\footnote{Note that we use the fact that the
lattice $\PV$ of projections in $V$ is complete. This is the main
reason why we chose von Neumann subalgebras rather than
$C^*$-algebras: the former contain enough projections, and their
projection lattices are complete.} Note that if $\P$ belongs to
$V$, then $\dasto{V}{P}=\P$. The mapping $\P\mapsto\dasto{V}{P}$
was originally introduced by de Groote in \cite{deG05},  who
called it the `$V$-support' of $\P$.

The key idea in this part of our  scheme is that rather than
thinking of a quantum proposition, $\SAin\De$,  as being
represented by a \emph{single} projection operator $\hat
E[A\in\De]$, we instead consider the \emph{collection}
$\{\delta\big(\hat E[A\in\De]\big)_V\mid V\in\Ob{\V{}}\}$ of  one
projection operator for each context $V$. As we will see, the link
with topos theory is that this collection of projectors is a
global element of a certain presheaf.

This `certain' presheaf is in fact the `outer' presheaf, which is
defined as follows:

{\definition The \emph{outer\footnote{In  the original papers by
CJI\ and collaborators, this was called the `coarse-graining'
presheaf, and was denoted $\ps{G}$. The reason for the change of
nomenclature will become apparent later.} presheaf} $\G$ is
defined over the category $\V{}$ as follows \cite{IB98,IB00}:
\begin{enumerate}
\item[(i)] On objects
$V\in\Ob{\V{}}$:  $\G_V:=\PV$

\item[(ii)] On morphisms $i_{V^{\prime}V}:V^{\prime
}\subseteq V:$ The mapping $\G(i_{V^{\prime} V}):\G_V
\map\G_{V^{\prime}}$ is given by
$\G(i_{V^{\prime}V})(\hat{\alpha}):=\dasto{V^\prime}{\alpha}$ for
all $\hat{\alpha}\in\PV$.
\end{enumerate}
}

With this definition, it is clear that the assignment $V\mapsto
\dasto{V}{P}$ defines a global element of the presheaf $\G$.
Indeed, for each context $V$, we have the projector
$\dasto{V}{P}\in\PV=\G_V$, and if $i_{V^{\prime}V}:V^{\prime
}\subseteq V$, then
\begin{equation}
\delta\big(\dasto{V}{P}\big)_{V^\prime}
=\bigwedge\big\{\hat{Q}\in\mathcal{P}(V^{\prime})\mid
\hat{Q}\succeq \dasto{V}{P}\big\}=\dasto{V^\prime}{P}
\end{equation}
and so the elements $\dasto{V}{P}$, $V\in\Ob{\V{}}$, are
compatible with the structure of the outer presheaf. Thus we have
a mapping
\begin{eqnarray}
\dasmap:\PH  &\map&\Ga\G                        \nonumber\\
\P  &  \mapsto&\{\dasto{V}{P}\mid {V\in\Ob{\V{}}\}}
\end{eqnarray}
from the projectors in $\PH$ to the global elements, $\Ga\G$, of
the outer presheaf.\footnote{Vis-a-vis our later work with the
language $\L{S}$, we should  emphasise that the outer presheaf has
no linguistic precursor:  in this sense, it has no fundamental
status. In fact, we could avoid it altogether and always work
directly with the spectral presheaf, $\Sig$---which, of course,
\emph{does} have a linguistic precursor. However, it is
technically useful to introduce the outer  presheaf as an
intermediate tool.}

\subsubsection{Properties of the Mapping $\dasmap:\PH\map\Ga\G$.}
\label{SubSubSec:PropDas} Let us now note some properties of the
map $\dasmap:\PH\map\Ga\G$ that are relevant to our overall
scheme.
\begin{enumerate}
\item For all contexts $V$, we have $\dasto{V}{0}=\hat0$, \ie\
the null projector  $\hat 0$ is mapped to the null element of
$\G$.

The null projector represents all propositions of the form
$\SAin\De$ with the property that $\sp{A}\cap\De=\varnothing$,
where $\sp{A}$ denotes the spectrum of the self-adjoint operator
$\A$. These propositions are trivially false.

\item For all contexts $V$, we have $\dasto{V}{1}=\hat 1$, \ie\
the unit operator $\hat 1$ is mapped to the `unit' element of
$\G$.

The unit operator $\hat 1$ represents all propositions of the form
$\SAin\De$  with the property that $\sp{A}\cap\De=\sp{A}$. These
propositions are trivially true.

\item There are global elements  of $\G$ that do \emph{not}
come from a projection operator. This will be discussed later.

However, if $\ga\in\G$ \emph{is} of the form $\das{P}$ for some
$\P$, then
\begin{equation}
                \P=\bigwedge_{V\in\Ob{\V{}}}\dasto{V}{P},
\end{equation}
because $\dasto{V}{P}\succeq \P$ for all $V\in\Ob{\V{}}$, and
$\dasto{V}{P}=\P$ for any  $V$ that contains $\P$.
\end{enumerate}

The next result is important as it means that no `information' is
lost in mapping a projection operator $\P$ to its associated
global element, $\das{P}$, of the presheaf $\G$.

\begin{theorem} The map $\dasmap:\PH\map\Ga\G$ is
injective.
\end{theorem}

To see this, let $\P_{1},\P_{2}$ be two different projection
operators in $\PH$. Without loss of generality, let
$\P_{1}\neq\hat 1 $. Let $V_{1}:=\{\P_{1} ,\hat
1\}^{\prime\prime}$ denote the abelian von Neumann algebra
generated by $\P_{1}$ and $\hat 1$. Then
$\mathcal{P}(V_{1})=\{\hat 0,\P_{1},\hat1-\P_{1},\hat 1\}$. We now
run through the possible relations between $\P_1$ and $\P_2$, case
by case.

Case (i): $\P_{1}\prec\P_{2}$. (Of course, this includes the case
$\P_{2}\prec\P_{1}$ by simple relabelling.) In this case,
$\dasto{V_{1}}{P_{1}}=\bigwedge \{\hat{Q}\in\mathcal{P}(V_{1})\mid
\hat{Q}\succeq \P_{1}\}=\P_{1}$ and $\dasto{V_{1}}{P_{2}}=\hat
1\neq \P_{1}$.                                          \\

Case (ii): $\hat 1-\P_{1}\prec\P_{2}$. Then
$\dasto{V_{1}}{P_{1}}=\P_{1}$ and
\begin{eqnarray}
\dasto{V_{1}}{P_{2}} &=&\bigwedge\big\{ \hat{Q}\in
\mathcal{P}(V_{1}) \mid\hat{Q}\succeq \P_{2}\big\}\nonumber\\
                    &\succeq&\bigwedge\big\{\hat{Q}\in
                    \mathcal{P}(V_{1})\mid\hat{Q}\succeq\hat 1-
                    \P_{1}\big\}\nonumber\\
                    & =&\hat 1-\P_{1}.
\end{eqnarray}

Case (iii): $\P_{1}=\hat 1-\P_{2}$. Then
$\dasto{V_{1}}{P_{1}}=\P_{1}$ and $\dasto{V_{1}}{P_{2}}=\hat
1-\P_{1}$.\\

Case (iv): $\P_{1}\not\prec\P_{2}$ and $\hat 1-\P_{1}\not\prec
\P_{2}$. In this case, $\dasto{V_{1}}{P_{1}}=\P_{1}$ and
\begin{eqnarray}
\dasto{V_{1}}{P_{2}} &=&\bigwedge\big\{\hat{Q}\in
\mathcal{P}(V_{1})\mid\hat{Q}\succeq \P_{2}\big\}\nonumber\\
                    & =&\hat 1.
\end{eqnarray}

In each case, $\dasto{V_{1}}{P_{1}}=
\P_{1}\neq\dasto{V_{1}}{P_{2}}$, and hence the global elements
$\das{P_{1}}=\{\dasto{V}{P_1}\mid V\in\Ob{\V{}}\}$ and
$\das{P_{2}}=\{\dasto{V}{P_2}\mid V\in\Ob{\V{}}\}$ differ.

\subsubsection{A Logical Structure for $\Ga\G$?}
\label{SubSec:LogStructG} We have seen that the quantities
$\das{P}:=\{\dasto{V}{P}\mid V\in\Ob{\V{}}\}$, $\P\in\PH$, are
elements of $\Ga\G$, and if they are to represent  quantum
propositions, one might expect/hope that (i) these global elements
of $\G$ form a Heyting algebra; and (ii) this algebra is related
in some way to the Heyting algebra of sub-objects of $\Sig$. Let
us see how far we can go in this direction.

Our first remark is that any two global elements $\ga_{1},\ga_{2}$
of $\G$ can be compared at each stage $V$ in the sense of logical
implication. More precisely, let $\ga_{1}(V)\in\PV$ denote the
$V$'th `component' of $\ga_1$, and ditto for $\ga_{2}(V)$. Then we
have the following result:
\begin{definition}
A partial ordering on $\Ga\G$ can be constructed in a `local' way
(\ie\ `local' with respect to the objects in the category $\V{}$)
by defining
\begin{equation}
\ga_{1}\succeq\ga_{2} \mbox{  if, and only if, } \forall
V\in\Ob{\V{}},\ \ga_{1}(V)\succeq\ga_{2}(V) \label{Def:g1>g2}
\end{equation}
where the ordering on the  right hand side of \eq{Def:g1>g2} is
the usual ordering in the lattice of projectors $\PV$.
\end{definition}
It is trivial to check that \eq{Def:g1>g2} defines a partial
ordering on $\Ga\G$. Thus $\Ga\G$ is a partially ordered set.

Note that if $\P,\hat Q$ are projection operators, then it follows
from \eq{Def:g1>g2} that
\begin{equation}
\das{P}\succeq\das{Q} \mbox{ if and only if } \P\succeq\hat Q
\end{equation}
since $\P\succeq\hat Q$ implies $\dasto{V}{P}\geq \dasto{V}{Q}$
for all contexts $V$.\footnote{On the other hand, in general,
$\P\succ \hat Q$ does not imply $\dasto{V}{P}\succ \dasto{V}{Q}$
but only $\dasto{V}{P}\succeq \dasto{V}{Q}$.} Thus the mapping
$\delta:\PH\map\Ga\G$ preserves the partial order.

The next thing is to see if a  logical $`\lor$'-operation can be
defined on $\Ga\G$. Once again, we try a `local' definition:
\begin{theorem}
A `$\lor$'-structure on $\Ga\G$ can be defined locally by
\end{theorem}
\begin{equation}
(\ga_1\lor\ga_2)(V):=\ga_1(V)\lor\ga_2(V)
        \label{Def:g1lorg2}
\end{equation}
for all $\ga_1,\ga_2\in\Ga\G$, and for all $V\in\Ob{\V{}}$.

\begin{proof}
It is not instantly clear that \eq{Def:g1lorg2} defines a global
element of $\G$. However, a key result in this direction is the
following:
\begin{lemma}
For each context $V$, and for all $\hat\alpha,\hat\beta\in\PV$, we
have
\begin{equation}
\G(i_{V^{\prime}V})(\hat{\alpha}\lor\hat\beta)=
\G(i_{V^{\prime}V})(\hat{\alpha})\lor
\G(i_{V^{\prime}V})(\hat{\beta}) \label{G(alorb)=}
\end{equation}
for all contexts $V^\prime$ such that  $V^\prime\subseteq V$.
\end{lemma}
The proof  is a straightforward consequence of the definition of
the presheaf $\G$.

One immediate consequence is that \eq{Def:g1lorg2} defines a
global element\footnote{The existence of the $\lor$-operation on
$\Ga\G$ can be extended to $\G$ itself. More precisely, there is
an arrow $\lor:\G\times \G\map\G$ where $\G\times\G$ denotes the
product presheaf over $\V{}$, whose objects are
$(\G\times\G)_V:=\G_V\times\G_V$. Then the arrow $\lor:\G\times
\G\map\G$ is defined at any context $V$ by
$\lor_{V}(\hat\alpha,\hat\beta):=\hat\alpha\lor\hat\beta$ for all
$\hat\alpha,\hat\beta\in\G_V$.} of $\G$. Hence the theorem is
proved.\end{proof}

It is also straightforward to  show that, for any pair of
projectors $\P,\hat Q\in\PH$, we have $\delta(\P\lor \hat
Q)_{V}=\dasto{V}{P}\vee\dasto{V}{Q}$, for all contexts
$V\in\Ob{\V{}}$. This means that, as elements of $\Ga\G$,
\begin{equation}
\delta(\P\vee \hat Q)=\das{P}\vee\das{Q}.
\end{equation}
Thus the mapping $\delta:\PH\map\Ga\G$ preserves the logical
`$\lor$' operation.

However, there is no analogous equation for the logical
`$\land$'-operation. The obvious local definition would be, for
each context $V$,
\begin{equation}
(\ga_1\land\ga_2)(V):=\ga_1(V)\land\ga_2(V)\label{Def:g1landg2}
\end{equation}
but this does not define a global element of $\G$ since, unlike
\eq{G(alorb)=}, for the $\land$-operation we have only
\begin{equation}
\G(i_{V^{\prime}V})(\hat{\alpha}\land\hat\beta)\;\preceq\;
\G(i_{V^{\prime}V})(\hat{\alpha})\land
\G(i_{V^{\prime}V})(\hat{\beta}) \label{G(alandb)=}
\end{equation}
for all $V^\prime\subseteq V$. As a consequence, for all $V$, we
have only the inequality
\begin{equation}
\delta(\P\land\hat Q)_{V}\;\preceq\;\dasto{V}{P}\land\dasto{V}{Q}
\label{delta(PlandQ)}
\end{equation}
and hence
\begin{equation}
\delta(\P\land\hat Q)\;\preceq\;\das{P}\land\das{Q}.
\end{equation}

It is easy to find examples where the inequality is strict. For
example, let $\P\neq \hat 0,\hat 1$ and $\hat Q=\hat 1-\P$. Then
$\P\land \hat Q=0$ and hence $\delta_{V}(\P\land \hat Q)=\hat 0$,
while $\dasto{V}{P}\land\dasto{V}{Q}$ can be strictly larger than
$\hat 0$, since $\dasto{V}{P}\succeq \P$ and $\dasto{V}{Q}\succeq
\hat Q$.

\subsubsection{Hyper-Elements of $\Ga\G$.}
We have seen that the global elements of $\G$, \ie\ the elements
of $\Ga\G$, can be equipped with a partial-ordering and a
`$\lor$'-operation, but attempts to define a `$\land$'-operation
in the same way fail because of the inequality in
\eq{delta(PlandQ)}.

However, the form of \eqs{G(alandb)=}{delta(PlandQ)} suggests the
following procedure. Let us define a \emph{hyper-element} of $\G$
to be an association, for each stage $V\in\Ob{\V{}}$, of an
element $\ga(V)\in\G_V$ with the property that
\begin{equation}
         \ga(V^\prime)\;\succeq\; \G(i_{V^\prime V})(\ga(V))
         \label{gammaVpsucceqG}
\end{equation}
for all $V^\prime\subseteq V$. Clearly every element of $\Ga\G$ is
a hyper-element, but not conversely.

Now, if $\ga_1$ and $\ga_2$ are hyper-elements, we can define the
operations `$\lor$' and  `$\land$' locally as:
\begin{eqnarray}
   (\ga_1\lor\ga_2)(V)&:=&\ga_1(V)\lor\ga_2(V)\\[2pt]
   (\ga_1\land\ga_2)(V)&:=&\ga_1(V)\land\ga_2(V)
\end{eqnarray}

Because of \eq{G(alandb)=} we have, for all $V^\prime\subseteq V$,
\begin{eqnarray}
\G(i_{V^{\prime}V})\big((\ga_1\land\ga_2)(V)\big)
     &=&\G(i_{V^{\prime}V})\big(\ga_1(V)\land \ga_2(V)\big)\\
        &\preceq&\G(i_{V^{\prime}V})(\ga_1(V))\land
                               \G(i_{V^{\prime}V})(\ga_2(V))\\
      &\preceq&\ga_1(V^{\prime})\land\ga_2(V^{\prime})\\
        &=&(\ga_1\land\ga_2)(V^{\prime})
\end{eqnarray}
so that the hyper-element condition \eq{gammaVpsucceqG} is
preserved.

The occurrence of a logical `$\lor$' and $\land$' structure is
encouraging, but it is not yet what we want. For one thing, there
is no mention of a negation operation; and, anyway, this is not
the expected algebra of sub-objects of a `state space' object. To
proceed further we must study more carefully the sub-objects of
the spectral presheaf.

\subsection{Daseinisation}
\subsubsection{From Global Sections of $\G$ to Sub-Objects
of $\Sig$.} The spectral presheaf, $\Sig$, played a central role
in the earlier discussions of quantum theory from a topos
perspective \cite{IB98,IB99,IB00,IB02}.
{\definition\label{Def_SpectralPresheaf} The \emph{spectral
presheaf}, $\Sig$, is defined as the following functor from
$\V{}^\op$ to $\Set$:
\begin{enumerate}
\item On objects $V$:  $\Sig_V$ is the Gel'fand spectrum of the unital, abelian
subalgebra $V$ of $\BH$; \ie\  the set of all multiplicative
linear functionals $\l:V\map\mathC$ such that $\l(\hat 1)=1$.

\item On morphisms $i_{V^{\prime}V}:V^\prime\subseteq V$:
$\Sig(i_{V^{\prime}V}):\Sig_V\map \Sig_{V^\prime}$ is defined by
$\Sig(i_{V^{\prime}V})(\l):= \l|_{V^\prime}$; \ie\ the restriction
of the functional $\l:V\map\mathC$ to the subalgebra
$V^\prime\subseteq V$.
\end{enumerate}
} \noindent One important result of spectral theory is that
$\Sig_V$ has a topology that is compact and Hausdorff, and with
respect to which the functionals are continuous. This will be
important in what follows \cite{KR83a}.

The spectral presheaf plays a fundamental role in our research
programme as applied to quantum theory. For example, it was shown
in the earlier work that the Kochen-Specker theorem \cite{KS67} is
equivalent to the statement that $\Sig$ has no global elements.
However, $\Sig$ \emph{does} have sub-objects, and these are
central to our scheme: {\definition A \emph{sub-object} $\ps{S}$
of the spectral presheaf $\Sig$ is a functor $\ps{S}:\V{}
^{op}\map\Set$ such that
\begin{enumerate}
\item$\ps{S}_V$ is a subset of $\Sig_V$ for all $V$.

\item  If $V^{\prime}\subseteq V$, then
$\ps{S}(i_{V^{\prime}V}):\ps{S}_V\map \ps{S}_{V^{\prime}}$ is just
the restriction $\l\mapsto\l|_{V^{\prime}}$ (\ie\ the same as for
$\Sig$), applied to the elements $\l\in \ps{S}_V\subseteq\Sig_V$.
\end{enumerate}
}

This definition of a sub-object is standard. However, for our
purposes we need something slightly different, namely  a `clopen'
sub-object. This is defined to be a sub-object $\ps{S}$ of $\Sig$
such that, for all  $V$, the set $\ps{S}_V$ is a
\emph{clopen}\footnote{A `clopen' subset of a topological space is
one that is both open and closed.} subset of the compact,
Hausdorff space $\Sig_V$. We denote by $\Subcl\Sig$ the set of all
clopen sub-objects of $\Sig$. We will show later that, like
$\Sub\Sig$, the set $\Subcl\Sig$ is a Heyting algebra.

This interest in clopen sets is easy to explain. For, according to
the Gel'fand spectral theory, a projection operator
$\hat\alpha\in\mathcal{P}(V)$  corresponds to a unique clopen
subset of the Gelfand spectrum, $\Sig_V$. Furthermore, the Gelfand
transform $\GT\alpha:\Sig_V\map\mathC$ of $\hat\alpha$ takes the
values $0,1$ only, since the spectrum of a projection operator is
just $\{0,1\}$.

It follows that $\GT{\alpha}$ is the characteristic function of
the subset, $S_{\hat\alpha}$, of $\Sig_V$, defined by
\begin{equation}
S_{\hat\alpha}:=\{\l\in\Sig_V \mid\l(\hat\alpha)=1\}.
\label{Def:Salphahat}
\end{equation}
The clopen nature of $S_{\hat\alpha}$  follows from the fact that,
by the spectral theory, the function
$\GT{\alpha}:\Sig_V\map\{0,1\}$ is continuous.

In fact, there is a lattice isomorphism between the lattice $\PV$
of projectors in $V$ and the lattice $\mathcal{C}L(\Sig_{V})$ of
clopen subsets of $\Sig_{V}$,\footnote{ The lattice structure on
$\mathcal{C}L(\Sig_{V})$ is defined as follows: if $(U_i)_{i\in
I}$ is an arbitrary family of clopen subsets of $\Sig_{V}$, then
$\bigcup_{i\in I}U_i$ is the maximum, and the \emph{interior} of
$\bigcap_{i\in I}U_i$ is the minimum of the family. One must take
the interior since $\bigcap_{i\in I}U_i$ is closed, but not
necessarily open.} given by
\begin{equation}
    \hat{\alpha}\mapsto S_{\hat{\alpha}}:=
\{ \l\in\Sig_{V}\mid\l(\hat{\alpha})=1\}.
\label{LatticeIsomProjsAndClopenSubsets}
\end{equation}
Conversely, given a clopen subset $S\in\mathcal{C}L(\Sig_{V})$, we
get the corresponding projection $\hat{\alpha}$ as the (inverse
Gelfand transform of the) characteristic function of $S$. Hence,
each $S\in\mathcal{C}L(\Sig_{V})$ is of the form
$S=S_{\hat{\alpha}}$ for some $\hat{\alpha}\in\PV$.

Our claim is the following:
\begin{theorem}
For each projection operator $\P\in\PH$, the collection of subsets
\begin{equation}
S_{\P}:=\{S_{\dasto{V}{P}}\subseteq\Sig_V\mid V\in\Ob{\V{}}\}
\end{equation}
forms a (clopen) \emph{sub-object} of the spectral presheaf
$\Sig$.
\end{theorem}
\begin{proof} To see this, let $\l\in S_{\dasto{V}{P}}$.
Then if $V^{\prime}$ is some abelian subalgebra of $V$, we have
$\dasto{V^{\prime}}{P}
=\bigwedge\big\{\hat{Q}\in\mathcal{P}(V^{\prime})\mid
\hat{Q}\succeq \dasto{V}{P}\big\}\succeq \dasto{V}{P}$. Now let
$\hat{Q}:=\dasto{V^{\prime}}{P}-\dasto{V}{P}$. Then
$\l\big(\dasto{V^{\prime}}{P}\big)
=\l\big(\dasto{V}{P}\big)+\l(\hat{Q})=1$, since
$\l\big(\dasto{V}{P}\big)=1$ and $\l(\hat{Q})\in\{0,1\}$. This
shows that
\begin{equation}
\{\l|_{V^{\prime}}\mid \l\in S_{\dasto{V}{P}}\}\subseteq
S_{\dasto{V^{\prime}}{P}}.\label{ldum1}
\end{equation}
However, the left hand side of \eq{ldum1} is the subset
$\G(i_{V^{\prime}V})(S_{\dasto{V}{P}})\subseteq\Sig_{V^\prime}$ of
the outer-presheaf restriction of elements in $S_{\dasto{V}{P}}$
to $\Sig_{V^{\prime}}$, and the restricted elements all lie in
$S_{\dasto{V^{\prime}}{P}}$. It follows that the collection of
sets
\begin{equation}
S_{\P}:=\{S_{\dasto{V}{P}}\subseteq\Sig_V\mid{V\in\Ob{\V{}}}\}
\end{equation}
forms a (clopen) sub-object of the spectral presheaf $\Sig$.
\end{proof}

By these means we have constructed a mapping
\begin{eqnarray}
\dasmap:\PH &\longrightarrow&\Subcl{\Sig}\nonumber\\
\P  &  \mapsto& S_{\P}:=\{S_{\dasto{V}{P}} \mid V\in\Ob{\V{}}\}
\label{delP(H)toSub}
\end{eqnarray}
which sends projection  operators on $\Hi$\ to clopen sub-objects
of $\Sig$.

\subsubsection{Heidegger encounters physics}
As usual, the projection $\P$ is regarded  as representing a
proposition about the quantum system. Thus $\delta$  maps
propositions about a quantum system to (clopen) sub-objects of the
spectral presheaf. This is remarkably analogous to the situation
in classical physics, in which propositions are represented by
subsets of the classical state space.

{\definition \noindent The map $\delta$ in \eq{delP(H)toSub} is a
fundamental part of our construction. We call it the
\emph{daseinisation}\footnote{The expression `daseinisation' comes
from the German word \emph{Dasein}, which plays a central role in
Heidegger's  existential philosophy. Dasein translates to
`existence' or, in the very literal sense often stressed by
Heidegger, to being-there-in-the-world (the hyphens are
\emph{very} important). Thus daseinisation
`brings-a-quantum-proposition-into-existence' (the hyphens are
\emph{very} important) by  hurling it into the classical
snap-shots of the world provided by the category of contexts. } of
$\P$. We shall  use the same word to refer to the operation in eq.
\ref{Def:dasouter} that relates to the outer presheaf.}

We will summarise here some useful properties of daseinisation.
\begin{enumerate}
    \item The null projection $\hat 0$ is mapped to the empty
     sub-object, $\delta(\hat 0)=\{\varnothing_{V}\mid
     V\in\Ob{\V{}}\}$, of $\Sig$.

    \item The identity projection $\hat 1$ is mapped to the
    unit sub-object, $\delta(\hat 1)=
    \{\Sig_V\mid V\in\Ob{\V{}}\}=\Sig$ of $\Sig$.

    \item Since the daseinisation map $\dasmap:\PH\map\Ga\G$
    is injective (see Section \ref{SubSubSec:PropDas}), and the
    mapping $\Ga\G\map\Ga(\PSig)$ is injective (because there
    is a monic arrow $\G\map\PSig$ in $\SetH{}$; see
    Section \ref{SubSec:MonicGPSig}), it follows that the
    daseinisation map
    $\dasmap:\PH\map\Ga(\PSig)\simeq\Subcl\Sig$
    is \emph{injective}. Thus no information about the projector
    $\P$ is lost when it is daseinised to become $\das{P}$.
\end{enumerate}

\subsection{The Heyting Algebra Structure on $\Subcl\Sig$}
The reason for daseinising projections is that the set,
$\Sub{\Sig}$, of sub-objects of the spectral presheaf forms a
\emph{Heyting algebra}. Thus the idea is to find a map
$\piqt:\PL{S}_0\map\Sub{\Sig}$ and then extend it to all of
$\PL{S}$ using the simple recursion ideas discussed in paper I
\cite{DI(1)}.

In our case, the act of daseinisation gives  a map from the
projection operators to the clopen sub-objects of $\Sub\Sig$, and
therefore  a map $\piqt:\PL{S}_0\map\Subcl{\Sig}$ can be defined
by
\begin{equation}
                \piqt(\Ain\De):=\delta\big(\hat E[A\in\De]\big)
\label{Def:pi(AinDelta)}
\end{equation}
However,  to extend this definition to $\PL{S}$, it is necessary
to show that the set of clopen sub-objects, $\Subcl{\Sig}$, is a
Heyting algebra. This is not completely obvious from the
definition alone.

\begin{theorem}{The collection, $\Subcl\Sig$, of all clopen
sub-objects of $\Sig$ is a Heyting algebra.}
\end{theorem}
\begin{proof} First recall how a Heyting algebra structure is placed on the set,
$\Sub{\Sig}$, of all sub-objects of $\Sig$.
\paragraph{The `$\lor$'- and `$\land$'-operations.}
Let $\ps{S},\ps{T}$ be two sub-objects of $\Sig$. Then the
`$\lor$' and `$\land$' operations are defined by
\begin{eqnarray}
(\ps{S}\lor\ps{T})_V & :=&\ps{S}_V\cup \ps{T}_V\\[2pt]
(\ps{S}\land \ps{T})_V& :=&\ps{S}_V\cap \ps{T}_V
\end{eqnarray}
for all contexts $V$. It is easy to see that if $\ps{S}$ and
$\ps{T}$ are clopen sub-objects of $\Sig$, then so are
$\ps{S}\lor\ps{T}$ and $\ps{S}\land \ps{T}$.

\paragraph{The zero and unit elements.}
The zero element in the Heyting algebra $\Sub{\Sig}$ is the empty
sub-object $\ps{0}:=\{\varnothing_{V}\mid V\in\Ob{\V{}}\}$, where
$\varnothing_{V}$ is the empty subset of $\Sig_V$. The unit
element in $\Sub{\Sig}$ is $\Sig$.  It is clear that both $\ps{0}$
and $\Sig$ are clopen sub-objects of $\Sig$.

\paragraph{The `$\Rightarrow$'-operation.}
The most interesting part is the definition of the implication
$\ps{S}\Rightarrow\ps{T}$. For all $V\in\Ob{\V{}}$, it is given by
\begin{eqnarray}
(\ps{S}\Rightarrow\ps{T})_V&:=&\{\l\in\Sig_V\mid\forall\,
V^{\prime}\subseteq V,\mbox{ if }
                                                   \nonumber\\
    &{}&\hspace{1cm} \Sig(i_{V^{\prime}V})(\l)\in
    \ps{S}_{V^\prime} \mbox{ then }\Sig(i_{V^{\prime}V})(\l)\in
            \ps{T}_{V^{\prime}}\}                       \\[2pt]
    &=&\{\l\in\Sig_V\mid \forall V^{\prime}
        \subseteq V,\mbox{ if}                \nonumber\\
    &{}&\hspace{1cm} \l|_{V^{\prime}}\in \ps{S}_{V^\prime}
        \mbox{ then }\l|_{V^{\prime}}\in\ps{T}_{V^{\prime}}\}.
\end{eqnarray}

Since $\lnot\ps{S}:=\ps{S}\Rightarrow\ps{0}$, the expression for
negation follows from the above as
\begin{eqnarray}
(\lnot \ps{S})_V & =&\{\l\in\Sig_V\mid\forall\,
        V^{\prime}\subseteq V,\
\Sig(i_{V^{\prime}V})(\l)\notin \ps{S}_{V^{\prime}}\}\\[5pt]
        & =&\{\l\in\Sig_V\mid\forall\,V^{\prime}\subseteq V, \,
              \l|_{V^{\prime}}\notin\ps{S}_{V^{\prime}}\}.
\end{eqnarray}
We rewrite the formula for negation as
\begin{equation}
(\lnot \ps{S})_V=\bigcap_{V^{\prime}\subseteq
V}\big\{\l\in\Sig_V\mid \l|_{V^{\prime}}\in
\ps{S}_{V^{\prime}}^c\big\}           \label{negS(V)=}
\end{equation}
where $\ps{S}_{V^{\prime}}{}^c$ denotes the complement of
$\ps{S}_{V^{\prime}}$ in $\Sig_{V^{\prime}}$. Clearly,
$\ps{S}_{V^{\prime}}{}^c$ is clopen in $\Sig_{V^{\prime}}$ since
$\ps{S}_{V^{\prime}}$ is clopen. Since the restriction
$\Sig(i_{V^{\prime}V}):\Sig_V\map\Sig_{V^{\prime}}$ is continuous
and surjective\footnote{See proof of Theorem \ref{Theorem:
SG=SigS} below.}, it is easy to see that the inverse image
$\Sig(i_{V^{\prime}V})^{-1}(\ps{S}_{V^{\prime}}{}^{c})$ is clopen
in $\Sig_V$. Clearly,
\begin{equation}
\Sig(i_{V^{\prime}V}
)^{-1}(\ps{S}_{V^{\prime}}{}^{c})=\big\{\l\in\Sig_V\mid
\l|_{V^{\prime}}\in \ps{S}_{V^{\prime}}{}^{c}\big\}
\end{equation}
and so, from \eq{negS(V)=} we have
\begin{equation}
(\lnot \ps{S})_V=\bigcap_{V^{\prime}\subseteq V}
\Sig(i_{V^{\prime}V})^{-1}(\ps{S}_{V^{\prime}}{}^{c})
                                        \label{lnotS(V)=}
\end{equation}

The problem is that we want $(\lnot\ps{S})_V$ to be a
\emph{clopen} subset of $\Sig_V$. Now the right hand side of
\eq{lnotS(V)=} is the intersection of a family, parameterised by
$\{V^\prime\mid V^\prime\subseteq V\},$ of clopen sets. Such an
intersection is always closed, but it is only guaranteed to be
open if $\{V^\prime\mid V^\prime\subseteq V\}$ is a finite set,
which of course may  not be the case.

If $V^{\prime\prime}\subseteq V^{\prime}$ and
$\l|_{V^{\prime\prime}}\in \ps{S}_{V^{\prime\prime}}{}^{c}$, then
$\l|_{V^{\prime}}\in \ps{S}_{V^{\prime}}{}^{c}$. Indeed, if we had
$\l|_{V^{\prime}}\in \ps{S}_{V^{\prime}}$, then
$(\l|_{V^{\prime}})|_{V^{\prime\prime}}= \l|_{V^{\prime\prime}}\in
\ps{S}_{V^{\prime}}$ by the definition of a sub-object, so we
would have a contradiction. This implies
$\Sig(i_{V^{\prime\prime}V}
)^{-1}(\ps{S}_{V^{\prime\prime}}{}^{c})\subseteq
\Sig(i_{V^{\prime}V})^{-1}(S_{V^{\prime}}{}^{c})$, and hence the
right hand side of \eq{lnotS(V)=} is a decreasing net of clopen
subsets of $\Sig_V$ which  converges to something, which we take
as the subset of $\Sig_V$ that is to be $(\neg\ps{S})_V$.

Here we have used the fact that the set of clopen subsets of
$\Sig_V$ is a complete lattice, where the minimum of a family
$(U_i)_{i\in I}$ of clopen subsets is defined as the interior of
$\bigcap_{i\in I}U_i$.  This leads us to define
\begin{eqnarray}
(\lnot \ps{S})_V  & :=&\mbox{int}\bigcap_{V^{\prime}\subseteq V}
\Sig(i_{V^{\prime}V} )^{-1}(\ps{S}_{V^{\prime}}{}^{c}) \\
    & =&\mbox{int}\bigcap_{V^{\prime}\subseteq V}\big\{\l\in
\Sig_V\mid \l|_{V^{\prime}}\in(S_{V^{\prime}}{}^{c})\big\}
\end{eqnarray}
as the negation in $\Subcl\Sig$. This modified definition
guarantees that $\lnot\ps{S}$ is a \emph{clopen} sub-object. A
straightforward extension of this method gives a consistent
definition of $\ps{S}\Rightarrow\ps{T}$.

\noindent This concludes the proof of the theorem.
\end{proof}

The important conclusion of this analysis is that $\Subcl\Sig$ is
a Heyting algebra. It is the  subset of $\Sub\Sig$ that
incorporates the spectral topology on the stalks $\Sig_V$, and the
spectral-theory association of projection operators in $V$ with
clopen subsets of $\Sig_V$. In particular, the  map
$\piqt:\PL{S}_0\map\Subcl{\Sig}$ given by
\begin{equation}
\piqt(\Ain\De):=\delta\big(\hat E[A\in\De]\big)
\label{Def:rho(AinDelta)(2)}
\end{equation}
can now be extended to all of $\PL{S}$ using the ideas discussed
in the first paper.

In conclusion: daseinisation can be used to give a
representation/model of the language $\PL{S}$ in the Heyting
algebra $\Subcl\Sig$.

\subsection{Daseinisation and the Operations of Quantum Logic.}
It is interesting to ask to what extent, if any, the map
$\delta:\PH\map\Subcl\Sig$ respects the lattice structure on
$\PH$. Of course, we know that it cannot be \emph{completely}
preserved since the quantum logic $\PH$ is non-distributive,
whereas $\Subcl\Sig$ is a Heyting algebra, and hence distributive.

We saw in Section \ref{SubSec:LogStructG} that, for the mapping
$\delta:\PH\map\Ga\G$, we have
\begin{eqnarray}
  \delta(\P\lor\hat{Q})_{V}&=&\dasto{V}{P}\lor\dasto{V}{Q},\\
 \delta(\P\land\hat{Q})_{V}&\preceq&\dasto{V}{P}\land\dasto{V}{Q}
\end{eqnarray}
for all contexts $V$ in $\Ob{\V{}}$.

The clopen subset of $\Sig_V$ that corresponds  to
$\dasto{V}{P}\lor\dasto{V}{Q}$ is $S_{\dasto{V}{P}}\cup
S_{\dasto{V}{Q}}$. This implies that the daseinisation map
$\delta:\PH\map\Subcl\Sig$ is a morphism of $\lor$-semi-lattices.

On the other hand, $\dasto{V}{P}\land\dasto{V}{Q}$ corresponds to
the subset $S_{\dasto{V}{P}}\cap S_{\dasto{V}{Q}}$ of $\Sig_V$.
Therefore, since $S_{\delta(\P\wedge \hat{Q})_{V}}\subseteq
S_{\dasto{V}{P}}\cap S_{\dasto{V}{Q}}$, daseinisation is
\emph{not} a morphism of $\wedge$-semi-lattices. In summary, for
all projectors $\P,\hat Q$ we have
\begin{eqnarray}
\delta(\P\lor\hat Q)&=&\delta(\P)\lor\delta(\hat Q)
                                            \label{del(PorQ)}\\[3pt]
\delta(\P\land\hat Q)&\preceq&\delta(\P)\land\delta(\hat Q)
                                            \label{del(PandQ)}
\end{eqnarray}
where the logical connectives on the left hand side lie in the
quantum logic $\PH$, and those on the right hand side lie in the
Heyting algebra $\Subcl\Sig$, as do the symbols `$=$' and
`$\preceq$'.

As remarked above, it is not surprising that  \eq{del(PandQ)} is
not an equality. Indeed, the quantum logic $\PH$ is
non-distributive, whereas the Heyting algebra $\Subcl\Sig$
\emph{is} distributive, and so it would be impossible for both
\eq{del(PorQ)} and \eq{del(PandQ)} to be equalities. The
inequality in \eq{del(PandQ)} is the price that must be paid for
liberating the projection operators from the shackles of quantum
logic and thrusting them down to the existential world of Heyting
algebras.

\subsubsection{The Status of the Possible Axiom
`$\Ain{\De_1}\land \Ain{\De_2}\Leftrightarrow
                     \Ain{\De_1\cap\De_2}$'}
We have the representation in \eq{Def:pi(AinDelta)},
$\piqt(\Ain\De):=\delta\big(\hat E[A\in\De]\big)$, of the
primitive propositions $\Ain\De$, and, as explained in our first
paper, \cite{DI(1)}, this can be extended to compound sentences by
making the obvious definitions:
\begin{eqnarray}
&(a)&\ \piqt(\alpha\lor\beta):=\piqt(\alpha)\lor\piqt(\beta)
                                                \label{pi(a)}\\
&(b)&\ \piqt(\alpha\land\beta):=\piqt(\alpha)\land\piqt\beta)
                                                \label{pi(b)}\\
&(c)&\ \piqt(\neg\alpha):=\neg\piqt(\alpha)\hspace{3cm}
                                                \label{pi(c)}\\
&(d)&\ \piqt(\alpha\Rightarrow\beta):=
        \piqt(\alpha)\Rightarrow\piqt(\beta)    \label{pi(d)}
\end{eqnarray}

As a result, we necessarily get a representation of the full
language $\PL{S}$ in the Heyting algebra $\Subcl\Sig$. However, we
then find that:
\begin{eqnarray}
        \piqt(\Ain{\De_1}\land\Ain{\De_2})&:=&
 \piqt(\Ain{\De_1})\land\piqt(\Ain{\De_2})\label{D1andD2a} \\
&=&\delta(\hat E[A\in\De_1])\land\delta(\hat E[A\in\De_2])
                                        \label{D1andD2b}\\
&\succeq& \delta(\hat E[A\in\De_1]\land \hat E[A\in\De_2])
                                         \label{D1andD2c}\\
&=&\delta(\hat E[A\in\De_1\cap\De_2)]) \label{D1andD2d}\\
&=&\piqt(\Ain{\De_1\cap\De_2})
\end{eqnarray}
where, \eq{D1andD2c} comes from \eq{del(PandQ)}, and in
\eq{D1andD2d} we have used the property of spectral projectors
that $\hat E[A\in\De_1]\land \hat E[A\in\De_2]= \hat
E[A\in\De_1\cap\De_2)]$. Thus, although by definition,
$\piqt(\Ain{\De_1}\land\Ain{\De_2})=
          \piqt(\Ain{\De_1})\land\piqt(\Ain{\De_2})$,
we only have the inequality
\begin{equation}
\piqt(\Ain{\De_1\cap\De_2})\preceq
\piqt(\Ain{\De_1}\land\Ain{\De_2})
\end{equation}

On the other hand, the same line of argument shows that
\begin{equation}
  \piqt(\Ain{\De_1}\lor\Ain{\De_2})=
                \piqt(\Ain{\De_1\cup\De_2})
\end{equation}
Thus we could consistently add the axiom
\begin{equation}
        \Ain{\De_1}\lor \Ain{\De_2}\Leftrightarrow
                     \Ain{\De_1\cup\De_2}
\end{equation}
to the language $\PL{S}$, but not
\begin{equation}
        \Ain{\De_1}\land \Ain{\De_2}\Leftrightarrow
                     \Ain{\De_1\cap\De_2}
\end{equation}
Of, course, both axioms are consistent with the representation of
$\PL{S}$ in classical physics.

It should be emphasised that there is nothing wrong with this
result: indeed, as stated above, it is the necessary price to be
paid for forcing a non-distributive algebra to have a
`representation' in a Heyting algebra.

\subsubsection{Inner Daseinisation and $\delta(\neg\P)$.}
In the same spirit, one might ask about ``$\neg(\Ain{\De})$''. By
definition, as in \eq{pi(d)}, we have
$\piqt(\neg(\Ain{\De})):=\neg\piqt(\Ain\De) =\neg\delta\big(\hat
E[A\in\De]\big)$. However, the question then is how, if at all,
this is related  to $\delta(\hat E[A\in\mathR/\De])$, bearing in
mind the axiom
\begin{equation}
 \neg(\Ain\De) \Leftrightarrow \Ain{\mathR\backslash\De}
\end{equation}
that can be added to the classical representation of $\PL{S}$.
Thus something needs to be said about $\delta(\neg\P),$ where
$\neg\P=\hat 1-\P$ is the negation operation in the quantum logic
$\PH$.

In order to express $\delta(\neg P)$ in terms of $\delta(\P)$, we
need to introduce another operation: {\definition  The \emph{inner
daseinisation}, $\delta^i(\P)$, of $\P$ is defined for each
context $V$ as
\begin{equation}
\dastoi{V}{P}:=\bigvee\big\{\hat{Q}\in\mathcal{P}(V)\mid
\hat{Q}\preceq \P\big\}. \label{Def:dasinner}
\end{equation}
} This should be contrasted with the definition of outer
daseinisation in eq. \eq{Def:dasouter}.

Thus $\dastoi{V}{P}$ is the best approximation that can be made to
$\P$ by taking the `largest' projector in $V$ that implies $\P$.

As with the other daseinisation construction, this operation  was
first introduced by de Groote in \cite{deG05} where he called it
the \emph{core} of the projection operator $\P$. We prefer to use
the phrase `inner daseinisation', and then to refer to
\eq{Def:dasouter} as the `outer daseinisation' operation on $\P$.
The existing notation $\dasto{V}{P}$ will be replaced with
$\dastoo{V}{P}$ if there is any danger of confusing the two
daseinisation operations.

With the aid of inner daseinisation, a new presheaf, $\H$, can be
constructed as an exact analogue of the the outer presheaf, $\G$,
defined in Section \ref{SubSub:defdas}. Specifically:

{\definition The \emph{inner presheaf} $\H$ is defined over the
category $\V{}$ as follows:
\begin{enumerate}
\item[(i)] On objects $V\in\Ob{\V{}}$: We define
$\H_V:=\PV$

\item[(ii)] On morphisms $i_{V^{\prime}V}:V^{\prime }\subseteq
V$:  The mapping $\H(i_{V^{\prime} V}):\H_V \map\H_{V^{\prime}}$
is defined as
$\H(i_{V^{\prime}V})(\hat{\alpha}):=\dastoi{V}{\alpha}$ for all
$\hat\alpha\in\PV$.
 \end{enumerate}
} It is easy to see that the collection $\{ \dastoi{V}{P}\mid
V\in\Ob{\V{}}\}$ of projection operators given by
\eq{Def:dasinner} is a global element of $\H$.

It is also straightforward to show that
\begin{equation}
  \G(i_{V^\prime V})(\neg\hat \alpha)=
\neg\, \H(i_{V^{\prime} V})(\hat\alpha) \label{Gi(negalpha)=}
\end{equation}
for all projectors $\hat\alpha$ in $V$, and for all
$V^\prime\subseteq V$. It follows from \eq{Gi(negalpha)=} that
\begin{equation}
\delta^{o}(\neg \P)_V=\hat 1-\dastoi{V}{P} \label{domega=}
\end{equation}
for all projectors $\P$ and all contexts $V$.

It is clear from \eq{Gi(negalpha)=} that the negation operation on
projectors defines a map $\neg:\Ga\G\map\Ga\H$, $\ga\mapsto
\neg\ga$; \ie\ for all contexts $V$, we map
$\ga(V)\mapsto\neg\ga(V):=\hat 1-\ga(V)$. In fact, one can go
further than this and show that there is a natural transformation
between the two presheaves $\G$ and $\H$ which is  an
\emph{isomorphism} in the category $\SetH{}$; \ie\ $\G$ and $\H$
are isomorphic objects in the topos. This isomorphism means that,
in principle, we can always work with one presheaf only. We have
elected to use the outer presheaf $\G$ and, as a result, only the
outer daseinisation (which we call just `daseinisation')
operation.

\section{The Special Nature of Daseinised Projections}
\label{Sec:SpecNatureDasP}
\subsection{Daseinised Projections as Optimal Sub-Objects}
We have shown how  daseinisation leads to an interpretation/model
of the language $\PL{S}$ in the Heyting algebra $\Subcl\Sig$. The
primitive propositions $\SAin\De$, \ie\ the elements of
$\PL{S}_0$, are represented by the clopen sub-objects $\das{P}$,
where \mbox{$\P=\hat E[A\in\De]$} is the spectral projection of
$\hat A$ that corresponds to the Borel set $\De$.

We also saw that, in general, the `minimum' $\das{P}\land\das{Q}$
of two projection operators $\P$ and $\hat Q$, is not itself of
the form $\das{R}$ for any projector $\hat R$.  The same applies
to the negation $\lnot\das{P}$.

This raises the question of whether the sub-objects of $\Sig$ that
\emph{are} of the form $\das{P}$ can be characterised in a simple
way. Rather interestingly, the answer is `yes' as we will now see.

Let $V^{\prime},V\in\Ob{\V{}}$ be such that $V^{\prime}\subseteq
V$. As one might expect, there is a close connection between the
restriction $\G(i_{V^{\prime}V}):\G_V\map\G_{V^{\prime}}$,
$\dasto{V}{P}\mapsto \dasto{V^{\prime}}{P}$, of the outer
presheaf, and the restriction $\Sig(i_{V^{\prime}V}):
\Sig_V\map\Sig_{V^{\prime}}$, $\l\mapsto\l|_{V^{\prime}}$, of the
spectral presheaf. Indeed, if $\P\in\PH$ is a projection,  and
$S_{\dasto{V}{P}}\subseteq\Sig_V$  is  defined as in
\eq{Def:Salphahat}, we have the following result:
\begin{theorem}
\label{Theorem: SG=SigS}
\begin{equation}
S_{\G(i_{V^{\prime}V})(\dasto{V}{P})}= \Sig (i_{V^{\prime}
V})(S_{\dasto{V}{P}}).
\end{equation}
\end{theorem}

Before proving the theorem, we need a few preparations. For
typographical simplicity, for a given pair $V^\prime\subseteq V$
we define the map
\begin{eqnarray}
        r:P(\Sig_V)  & \map&
                    P(\Sig_{V^{\prime}})\nonumber\\
        S   & \mapsto&\Sig(i_{V^{\prime}V})(S)=\{\l
|_{V^{\prime}}\mid \l\in S\}, \label{Def:r}
\end{eqnarray}
where $P(\Sig_V)$ is the power set of $\ps{\Si}_V$. We then have

\begin{lemma}
\label{L_1}If $\dasto{V^{\prime}}{P}=\dasto{V}{P}$, then every
$\l_{V^{\prime}}\in S_{\dasto{V^{\prime}}{P}}$ is of the form
$\l_{V}|_{V^{\prime}}$ for some $\l_{V}\in S_{\dasto{V}{P}}$. This
implies $r(S_{\dasto{V}{P}})=S_{\dasto{V^{\prime}}{P}}$.
\end{lemma}

\begin{proof} If $\l_{V}\in S_{\dasto{V}{P}}$, then $\l_{V}
|_{V^{\prime}}\in S_{\dasto{V^{\prime}}{P}}$ (since $\l
_{V}|_{V^{\prime}}(\dasto{V^{\prime}}{P})=\l_{V}|_{V^{\prime}
}(\dasto{V}{P})=1$), whereas if $\l_{V}\notin S_{\dasto{V}{P}}$,
then $\l_{V}|_{V^{\prime}}\notin S_{\dasto{V^{\prime}}{P}}$ (since
$\l_{V}|_{V^{\prime}}(\dasto{V^{\prime}}{P})=
\l_{V}|_{V^{\prime}}(\dasto{V}{P})=0$). Since \emph{every}
$\l_{V^{\prime}}\in\Sig_{V^{\prime}}$ is of the form
$\l_{V}|_{V^{\prime}}=r(\l_{V})$ for some $\l_{V}\in
S_{\dasto{V}{P}}$, the mapping $r:P(\Sig_{V}) \map
P(\Sig_{V^{\prime}})$ sends $S_{\dasto{V}{P}}$ to
$S_{\dasto{V^{\prime}}{P}}$, and the complement
$(S_{\dasto{V}{P}})^{c}$ in $\Sig_{V}$ is sent to the complement
$(S_{\dasto{V^{\prime}}{P}})^{c}$ in $\Sig_{V^{\prime}}$.
\end{proof}

\begin{proof}\textbf{(of Theorem \ref{Theorem: SG=SigS})} By definition,
$\G(i_{V^{\prime}V})\big(\dasto{V}{P}\big)=\dasto{V^\prime}{P}$.
The mapping $r$ defined in \eq{Def:r} is open and continuous
(which follows from Prop. 3.22 in \cite{deG05c}\footnote{The von
Neumann algebra $V$ corresponds to the algebra $\mathcal{R}$ in de
Groote's (more general) proof, $V^{\prime}$ corresponds to
$\mathcal{A}$. Our mapping $r$ is his $\zeta_{\mathcal{A}}$. For
abelian von Neumann algebras, quasipoints can be identified with
elements of the Gel'fand spectrum, and the topologies on the space
$\mathcal{Q}(V)$ of quasipoints and the Gel'fand spectrum
$\Sig_{V}$ coincide. For all this, see \cite{deG05c}, in
particular, Thm. 3.2.}).

The map $r$ is also closed. To see this, let $C\subseteq\Sig_{V}$
be a closed subset. Since $\Sig_{V}$ is compact, $C$ is compact,
and since $r$ is continuous, $r(C)\subseteq\Sig_{V^{\prime}}$ is
compact, too. However, $\Sig_{V^{\prime}}$ is Hausdorff, and so
$r(C)$ is closed in $\Sig_{V^{\prime}}$.

Now, $S_{\dasto{V}{P}}$ is a clopen subset of $\Sig_{V}$, so
$r(S_{\dasto{V}{P}})$ is a clopen subset of $\Sig_{V^{\prime}}$.
Clearly, we have
\begin{eqnarray}
r(S_{\dasto{V}{P}}) &  =&{\rm int}\bigcap\{S\subseteq\Sig
(V^{\prime}) \mid S\in\mathcal{C}L(\Sig(V^{\prime
})),\ r(S_{\dasto{V}{P}})\subseteq S\}                  \\
&=&{\rm int}\bigcap\{S_{\hat{Q}}\in\mathcal{C}L(\Sig_{V^{\prime}})
\mid r(S_{\dasto{V}{P}})\subseteq S_{\hat{Q}}\}.
\end{eqnarray}

We now show that $r(S_{\dasto{V}{P}})\subseteq S_{\hat{Q}}$
implies $\hat{Q}\succeq\dasto{V}{P}$. Assume that $\hat
{Q}\prec\dasto{V}{P}$. Let $\hat{R}:=\dasto{V}{P}-\hat{Q}\in\PV$,
and let $\l\in S_{\hat{R} }\subseteq\Sig_{V}$. Then $\l\in
S_{\dasto{V}{P}}$, but $\l\notin S_{\hat{Q}}\subseteq\Sig_{V}$.
Lemma \ref{L_1} shows that $r(S_{\hat{Q}})=S_{\hat{Q}}$ (where the
former $S_{\hat{Q}}$ is a subset of $\Sig_{V}$ and the latter is a
subset of $\Sig_{V^{\prime}}$) and $r((S_{\hat{Q}}
)^{c})=(S_{\hat{Q}})^{c}$. In particular, $\l|_{V^{\prime}}\notin
S_{\hat{Q}}\subseteq\Sig_{V^{\prime}}$, but $\l |_{V^{\prime}}\in
r(S_{\dasto{V}{P}})$, so
\begin{equation}
\hat{Q}\prec\dasto{V}{P}\quad\Rightarrow\quad
r(S_{\dasto{V}{P}})\nsubseteq S_{\hat{Q}}.
\end{equation}
Hence we must have $\hat{Q}\succeq\dasto{V}{P}, $ from which we
obtain
\begin{equation}
r(S_{\dasto{V}{P}})={\rm int}\bigcap\{S_{\hat{Q}}
\in\mathcal{C}L(\Sig_{V^{\prime}})\mid\hat{Q}\succeq\dasto{V}{P}\}.
\end{equation}
We now use the lattice isomorphism between
$\mathcal{C}L(\Sig_{V^{\prime}})$ and $\mathcal{P}(V^{\prime})$
(see \eq{LatticeIsomProjsAndClopenSubsets}) to get
\begin{eqnarray}
r(S_{\dasto{V}{P}})  & =&{\rm int}\bigcap\{S_{\hat{Q}
}\in\mathcal{CL}(\Sig_{V^{\prime}})\mid\hat{Q}\succeq\dasto{V}{P}\}
                                                \nonumber\\
      & =&S_{\{\hat{Q}\in\mathcal{P}(V^{\prime})\mid
            \hat{Q}\succeq\dasto{V}{P}\}}
                  =S_{\G(i_{V^{\prime}V})(\dasto{V}{P})}.
\end{eqnarray}
This completes the proof of the theorem.
\end{proof}

This result shows that the sub-objects $\das{P}=
\{S_{\dasto{V}{P}}\mid V\in\Ob{\V{}}\}$ of $\Sig$ are of a very
special kind. Namely, they are such that the restriction
\begin{equation}
\Sig(i_{V^{\prime}V}):S_{\dasto{V}{P}}\map
                        S_{\dasto{V^\prime}{P}}
\end{equation}
is a \emph{surjective} mapping of sets.

For an arbitrary sub-object $\ps{K}$ of $\Sig$, this will
generally not be the case and $\Sig(i_{V^{\prime}V})$ will only
map $\ps{K}_V$ \emph{into} $\ps{K}_{V^\prime}$. Indeed, this is
essentially the \emph{definition} of a sub-object of a presheaf.
Thus we see that the daseinised projections $S_{\dasto{}{P}}=
\{S_{\dasto{V}{P}}\mid V\in\Ob{\V{}}\}$ are optimal in the
following sense. As we go `down the line' to smaller and smaller
subalgebras of a context $V$---for example, from $V$ to
$V^{\prime}\subseteq V$, then to $V^{\prime\prime }\subseteq
V^{\prime}$ etc.---then the subsets $S_{\dasto{V^\prime}{P}}$,
$S_{\dasto{V^{\prime\prime}}{P}}$,... are as small as they can be;
\ie\ $S_{\dasto{V^\prime}{P}}$ is the \emph{smallest} subset of
$\Sig_{V^{\prime}}$ such that
$\Sig(i_{V^{\prime}V})(S_{\dasto{V}{P}})\subseteq
S_{\dasto{V^\prime}{P}}$, likewise
$S_{\dasto{V^{\prime\prime}}{P}}$ is the smallest subset of
$\Sig_{V^{\prime\prime}}$ such that
$\Sig(i_{V^{\prime\prime}V^{\prime}})
(S_{\dasto{V^\prime}{P}})\subseteq
S_{\dasto{V^{\prime\prime}}{P}}$, and so on.

It is also clear from this result that there are sub-objects of
$\Sig$ that are \emph{not} of the form $\das{P}$ for any projector
$\P\in\PH$.

These more general sub-objects of $\Sig$ show up explicitly in the
representation of the more sophisticated language $\L{S}$. This
subject will be discussed in the next paper, III, in this series.
There we analyse the representation, $\phi$, of the language
$\L{S}$ in the topos $\SetH{}$. This involves constructing the
quantity-value object $\R_\phi$ (to be denoted $\ps{\R}$), and
then finding the representation of a function symbol
$A:\Si\map\R$, from $\L{S}$, in the form of a specific arrow
$\breve{A}:\Sig\map\ps{\R}$ in the topos. The generic sub-objects
of $\Sig$ are then of the form $\breve{A}^{-1}(\ps{\Xi})$ for
sub-objects $\ps{\Xi}$ of $\ps\R$. This is an illuminating way of
studying the sub-objects of $\Sig$ that do not come from the
propositional language $\PL{S}$.

\section{Truth Values in Topos Physics}
\label{Sec:TruthValues}
\subsection{The Mathematical Proposition ``$x\in K$''}
So far we have concentrated on finding a Heyting-algebra
representation of the propositions in quantum theory, but  of
course there is more to physics than that. We also want to know if
a certain proposition is \emph{true}: a question which, in
physical theories, can only be answered by specifying a
\emph{state} of the system, or something that can play an
analogous role.

In classical physics, the situation is straightforward. There, a
proposition $\SAin\De$ is represented by the subset\footnote{Here,
$\breve{A}:\S\map\mathR$ is the mathematical representation of the
physical quantity $A$.}
$\picl(\Ain\De):=\breve{A}^{-1}(\Delta)\subseteq\S$ of the state
space $\S$; and then, the proposition is true in a state $s$ if
and only if $s\in\breve{A}^{-1}(\De)$; \ie\ if and only if the
(micro-) state $s$ belongs to the subset, $\picl(\Ain\De)$, of
$\S$ that represents the proposition.

Thus, each state $s$ assigns to any primitive proposition
$\SAin\De$, a truth value, $\TVal{\Ain\De}{s}$,  which lies in the
set $\{{\rm false},{\rm true}\}$ (which we identify with
$\{0,1\}$) and is defined as\
\begin{equation}\label{Def:[AinD]Class}
        \TVal{\Ain\De}{s}:=
        \left\{\begin{array}{ll}
            1 & \mbox{\ if\ $s\in\picl(\Ain\De):=
                                \breve{A}^{-1}(\De)$;} \\
            0 & \mbox{\ otherwise.}
         \end{array}
        \right.
\end{equation}
for all $s\in\S$.

However, the situation is very different for quantum theory. This
is because the spectral presheaf $\Sig$, which is the analogue of
the classical state space $\S$, has no global elements at all. And
no doubt there will  be other topos theories of physics where
$\Ga\Si_\phi$ is empty; or, if $\Si_\phi$ does have global
elements, there are not enough of them to determine $\Si_\phi$ as
an object in the topos.  In this circumstance,  a new concept is
required to replace the familiar idea of a `state of the system'.

In physics, the propositions of interest are of the form
$\SAin\De$, which refer to the value of a physical quantity.
However, in constructing a theory of physics, physical
propositions are invariably first translated into
\emph{mathematical} propositions; the concept of `truth' is then
studied in the context of the latter.

Let us start with set-theory based mathematics, where the most
basic proposition is of the form ``$x\in K$'', where $K$ is a
subset of a set $X$, and $x$ is an element of $X$. Then the truth
value, denoted $\TValM{x\in K}$, of the proposition ``$x\in K$''
is
\begin{equation}
 \TValM{x\in K}=
        \left\{\begin{array}{ll}
            1 & \mbox{\ if\ $x$ belongs to $K$;} \\
            0 & \mbox{\ otherwise.}
         \end{array}
        \right.                                 \label{TVxinKcl}
\end{equation}
Thus the proposition ``$x\in K$'' is true if, and only if, $x$
 belongs to $K$; \ie\ if, and only if, $x\in K$.

This result sounds like a tautology but, nevertheless, it is the
foundation of the assignment of truth values in physics. For
example, in classical physics, if the state is $s$, the truth
value, $\TVal{\Ain\De}{s}$,  of the physical proposition
$\SAin\De$ is defined to be the truth value of the
\emph{mathematical} proposition ``$\breve{A}(s)\in\De$''; or,
equivalently, of the mathematical proposition ``$s\in
\breve{A}^{-1}(\De)$''.

Thus, using \eq{TVxinKcl}, we get, for all $s\in\S$,
\begin{equation}
        \TVal{\Ain\De}{s}:=
        \left\{\begin{array}{ll}
            1 & \mbox{\ if\ $s\in\breve{A}^{-1}(\De)$;}
         \\
            0 & \mbox{\ otherwise.}\label{Def:[AinD]Class2}
         \end{array}
        \right.
\end{equation}
which reproduces \eq{Def:[AinD]Class}. We now consider the
analogue of the above in a general topos $\tau$.

Let $X$ be an object in the topos $\tau$, and let $\so{K}$ be a
sub-object of $X$. Then $\so{K}$ is determined by a characteristic
arrow $\cha{K}: X \map\O_\tau$, where $\O_\tau$ is the sub-object
classifier for the topos; equivalently we have an arrow
$\name{K}:1_\tau\map PX$.

Now suppose that\footnote{One of the basic properties of a topos
is that there is a one-to-one correspondence between arrows
$f:A\times B\map\O$ and arrows $\name{f}:A\map PB:=\O^B$. In
general, $\name{f}$ is called the \emph{power transpose} of $f$.
If $A\simeq 1$ then  $\name{f}$ is known as the \emph{name} of the
arrow $f:B\map\O$. On exponentials, see the Appendix in paper I.}
$\name{x}:1_\tau\map X$ is a global element of $X$; \ie\ $x\in\Ga
X$. Then the truth value of the mathematical proposition
``$x\in\so{K}$'' is defined to be
\begin{equation}
        \TValM{x\in\so{K}}:=\chi_{\so{K}}\circ \name{x}
        \label{VxinpsK}
\end{equation}
where  $\cha{K}\circ \name{x}:1_\tau\map\O_\tau$. Thus
$\TValM{x\in\so{K}}$ is an element of $\Ga\O_\tau$; \ie\ it is a
global element of the sub-object classifier $\O_\tau$.

The relation with the result  \eq{TVxinKcl} (in the topos $\Set$)
can be seen by noting that, in \eq{TVxinKcl}, the characteristic
function of the subset $K\subseteq X$ is the function
$\cha{K}:X\map\{0,1\}$ such that $\chi_K(x)=1$ if $x\in K$, and
$\chi_K(x)=0$ otherwise. It follows that \eq{TVxinKcl} can be
rewritten as
\begin{eqnarray}
        \TValM{x\in K}&=&\chi_K(x) \nonumber         \\
                    &=&\chi_K\circ\name{x}\label{TVxinKcl(b)}
\end{eqnarray}
where, in \eq{TVxinKcl(b)}, $\name{x}$ is the function
$\name{x}:\{*\}\map X$ that is defined by $\name{x}(*):=x$. The
link with \eq{VxinpsK} is clear when one remembers that, in the
topos $\Set$, the terminal object, $1_\Set$, is just the singleton
set $\{*\}$.

In quantum theory,  the topos is $\SetH{}$, and so the objects are
all presheaves. In particular, at each stage $V$, the sub-object
classifier $\Om:=\O_{\SetH{}}$  is the set of sieves on $V$. In
this case, if $\ps{K}$ is a sub-object of $\ps{X}$, and
$x\in\Ga\ps{X}$, the explicit form for \eq{TVxinKcl(b)} is the
sieve
\begin{equation}
  \TValM{x\in\ps{K}}_V:=\{V^\prime\subseteq V
                \mid x_{V^\prime}\in \ps{K}_{V^\prime}\}
        \label{ValxinKpresheaf}
\end{equation}
at each stage $V\in\Ob{\V{}}$.

The definitions \eq{VxinpsK} and \eq{ValxinKpresheaf} play a
central role in constructing truth values in our scheme. However,
as $\Sig$ has no global elements, these truth values cannot be
derived from an expression $\TValM{s\in\ps{K}}$ with
$\name{s}:1_{\SetH{}}\map\Sig$. Therefore, we must proceed in a
different way, as will become clear by the end of the following
Section.

\subsection{Truth Objects}\label{SubSub:TruthObjects}
\subsubsection{Linguistic Aspects of Truth Objects.}
To understand the construction of `truth values' in a topos we
return again to the discussion in paper I of  local languages
\cite{DI(1)}. In this Section we will employ the local language
$\L{S}$, rather than the  propositional language, $\PL{S}$,  used
earlier in this paper.

Thus, let $\L{S}$ be the local language for a system $S$. As
explained in paper I, this is a typed language whose minimal set
of ground-type symbols is $\Si$ and $\R$. There is a non-empty
set, $\F{S}$, of function symbols $A:\Si\map\R$ that correspond to
the physical quantities of $S$.

Now  consider a  representation, $\phi$, of $\L{S}$ in a topos
$\tau_\phi$.  As discussed in paper I, the propositional aspects
of the language $\L{S}$ are captured in the term
$\q{A(\va{s})\in\va\De}$ of type $\O$, where $\va{s}$ and
$\va{\De}$ are  variables of type $\Si$ and $P\R$ respectively
\cite{DI(1)}. In  a topos representation, $\phi$, the
representation, $\Val{A(\va{s})\in\va\De}_\phi$, of the term
$\q{A(\va{s})\in\va\De}$  is given by the chain of
arrows\footnote{In \eq{A(s)intildeDeChain}, $e_{\R_\phi}:
\R_\phi\times P\R_\phi\map\O_{\tau_\phi}$ is the evaluation arrow
associated with the power object $P\R_\phi$.}\cite{Bell88}
\begin{equation}
\Si_\phi\times P\R_\phi\mapright{A_\phi\times\id}
        \R_\phi\times P\R_\phi\mapright{e_{\R_\phi}}\O_{\tau_\phi}
                                        \label{A(s)intildeDeChain}
\end{equation}
in the topos $\tau_\phi$. Then, if $\name{\Xi}:1_{\tau_\phi}\map
P\R_\phi$ is a sub-object of the quantity-value object $\R_\phi$,
we get the chain
\begin{equation}
\Si_\phi\simeq\Si_\phi\times 1_{\tau_\phi}\mapright{\id\times
\name{\Xi}} \Si_\phi\times P\R_\phi\mapright{A_\phi\times\id}
        \R_\phi\times P\R_\phi\mapright{e_{\R_\phi}}\O_{\tau_\phi}.
\end{equation}
which is interpreted as the characteristic arrow associated with
the proposition $\SAin \Xi$.

Equivalently,  we can use the term, $\{\va{s}\mid
A(\va{s})\in\va\De\}$, which has a free variable $\va\De$ of type
$P\R$ and is of type $P\Si$ (see paper I, \cite{DI(1)}). This term
is represented by the arrow $\Val{\{\va{s}\mid
A(\va{s})\in\va\De\}}_\phi : P\R_\phi\map P\Si_\phi$, which  is
the power transpose of
 $\Val{A(\va{s})\in\va\De}_\phi$:
\begin{equation}
\Val{\{\va{s}\mid A(\va{s})\in\va\De\}}_\phi =
\name{\Val{A(\va{s})\in\va\De}_\phi}\label{[]=nametildeDe}
\end{equation}
The proposition $\SAin\Xi$ is then represented by the arrow
$\Val{\{\va{s}\mid
A(\va{s})\in\va\De\}}_\phi\circ\name{\Xi}:1_{\tau_\phi}\map
P\R_\phi$.

We note an important difference with the analogous situation for
the language $\PL{S}$. In propositions of the type $\SAin\De$, the
symbol `$\Delta$' is \emph{external} to the language (it is a
specific subset of $\mathR$), and it is \emph{independent} of the
representation of $\PL{S}$. However, in the case of $\L{S}$, the
variable $\va{\De}$ is \emph{internal} to the language, and the
quantity $\Xi$ in the proposition $\SAin \Xi$ is a sub-object of
$\R_\phi$ in a \emph{specific} topos representation,  $\phi$, of
$\L{S}$.

This is how physical propositions are represented mathematically.
But  how are truth values to be assigned to these propositions? In
the topos $\tau_\phi$ a truth value is an element of the Heyting
algebra $\Ga\O_{\tau_\phi}$. Thus the challenge is to assign a
global element of $\O_{\tau_\phi}$ to each proposition associated
with the representation of the term  $\{\va{s}\mid
A(\va{s})\in\va\De\}$ of type $P\Si$;  (or, equivalently, the
representation of the term `$A(\va{s})\in\va\De$').

Let us first pose this question at a linguistic level. In a
representation $\phi$, an element of $\Ga\O_{\tau_\phi}$ is
associated with a representation of a term of type $\O$ with no
free variables. Hence the question can be rephrased as asking how
a term, $t$, in $\L{S}$ of type $P\Si$ can be `converted' into a
term of type $\O$? At this stage, we are happy to have free
variables in this term, in which case it will be represented by an
arrow in $\tau_\phi$ with co-domain $\O_{\tau_\phi}$, but whose
domain is other than $1_{\tau_\phi}$. This is an intermediate
stage to obtaining a global element of $\O_{\tau_\phi}$.

In the context of the language $\L{S}$ there are two obvious ways
of `converting' the term $t$ of type $P\Si$ to a term of type
$\O$:
\begin{enumerate}
        \item Choose a term, $s$, of type $\Si$; then  the
        term `$s\in t$' is of type $\O$.

        \item Choose a term,  $\TO$, of type $PP\Si$; then
        the term `$t\in \TO$' is of type $\O$.
\end{enumerate}

In regard to the first option, the simplest term of type $\Si$ is
a variable $\va{s_1}$ of type $\Si$. Then, the term `$\va{s_1}\in
\{\va{s}\mid A(\va{s})\in\va{\De}\}$' is of type $\O$ with the
free variables $\va{s_1}$ of type $\Si$ and $\va{\De}$ of type
$P\R$. However, the axiom of comprehension in $\L{S}$ says that
\begin{equation}
\va{s_1}\in \{\va{s}\mid A(\va{s})\in\va{\De}\}\Leftrightarrow
        A(\va{s_1})\in\va{\De}
\end{equation}
and so we are back with the term `$A(\va{s})\in\va{\De}$', which
is of type $\O$ and with the free variable $\va{s}$ of type $\Si$.

As stated above,  the $\phi$-representation,
$\Val{A(\va{s})\in\va\De}_\phi$, of $\q{A(\va{s})\in\va{\De}}$ is
the chain of arrows in \eq{A(s)intildeDeChain}. Now, if the
representation, $\phi$, is such that there exist global elements,
$\name{s}:1_{\tau_\phi}\map\Si_\phi$, of $\Si_\phi$, then each
such element can be regarded as a `(micro)-state' of the system.
This can be combined with the arrow
$\Val{A(\va{s})\in\va{\De}}_\phi:\Si_\phi\times
P\R_\phi\map\O_{\tau_\phi}$ to give the arrow
\begin{equation}
\Val{A(\va{s})\in\va{\De}}_\phi\circ (\name{s}\times \id):
1_{\tau_\phi}\times P\R_\phi\map
      \Si_\phi\times  P\R_\phi\map\O_{\tau_\phi} \label{AinDelcircs0}
\end{equation}
Finally, if $\name{\Xi}:1_{\tau_\phi}\map P\R_\phi$ is a
sub-object of the quantity-value object $\R_\phi$, then
\begin{equation}
\Val{A(\va{s})\in\va{\De}}_\phi\circ (\name{s}\times
\id)\circ\name{\Xi}:
1_{\tau_\phi}\map\O_{\tau_\phi}\label{AinDelcircs2}
\end{equation}
is the desired global element of $\O_{\tau_\phi}$.  To simplify
the notation somewhat, we will rewrite \eq{AinDelcircs2} as
\begin{equation}
\Val{A(\va{s})\in\va{\De}}_\phi\la \name{s},\name{\Xi}\ra:
1_{\tau_\phi}\map\O_{\tau_\phi}\label{AinDelcircs}
\end{equation}
Thus, $\la\name{s},\name{\Xi}\ra:1_{\tau_\phi}\map\Si_\phi\times
P\R_\phi$.

In other words, when the `state of the system' is
$s\in\Ga\Si_\phi$, the `truth value' of the proposition $\SAin
\Xi$---as represented by the sub-object of $\Si_\phi$ with
characteristic arrow
$\Val{A(\va{s})\in\va{\De}}_\phi\circ(\id\circ\name{\Xi}):
\Si_\phi\times 1\map\O_{\tau_\phi}$---is the global element of
$\O_{\tau_\phi}$ defined by $\Val{A(\va{s})\in\va{\De}}_\phi\la
\name{s},\name{\Xi}\ra: 1_{\tau_\phi}\map\O_{\tau_\phi}$. All this
may seem rather complicated but, in fact, it is quite
straightforward once one gets used to the notation.

This is the procedure that is adopted in classical physics, where
a truth value is assigned to propositions by specifying a
micro-state, $s\in\Si_\s$, where $\Si_\s$ is the classical state
space in the representation $\s$ of $\L{S}$. Specifically, for all
$s\in\Si_\s$, the truth value of the proposition $\SAin\De$ as
given by \eq{AinDelcircs}, is (c.f. \eq{Def:[AinD]Class})
\begin{equation}
\TVal{\Ain\De}{s}
        =\Val{A(\va{s})\in\va{\De}}_\s(s,\De)=
        \left\{\begin{array}{ll}
            1 & \mbox{\ if\ $A_\s(s)\in\De$;}\\
            0 & \mbox{\ otherwise.}
         \end{array}
        \right.                 \label{nuAinDe;s}
\end{equation}
where $\Val{A(\va{s})\in\va{\De}}_\s:\Si_\s\times
P\mathR\map\O_{\tau_\s}\simeq\{0,1\}$. Thus we recover the earlier
result \eq{Def:[AinD]Class2}.

\subsubsection{Truth Objects in a General Topos.}
By hindsight, we know that the option to use global elements of
$\Si_\phi$ is not available in the quantum case. There, the state
object, $\Sig$, is the spectral presheaf, and this has no global
elements by virtue of the Kochen-Specker theorem. The absence of
global elements of the state object $\Si_\phi$ could  well be true
in many other topos models of physics, and therefore an
alternative general strategy is needed to that employing
micro-states $\name{s}:1_{\tau_\phi}\map\Si_\phi$.

This takes us to the second possibility: namely, to introduce a
term, $\TO$,    of type $PP\Si$, and then work with the term
`$\{\va{s}\mid A(\va{s})\in\va{\De}\} \in\TO$', which is  of type
$\O$, and has whatever free variables are contained in $\TO$, plus
the variable $\va{\De}$ of type $P\R$.

The simplest choice is to let the term of type $PP\Si$ be a
variable, $\va{\TO}$, of type $PP\Si$, in which case the term
$\q{\{\va{s}\mid A(\va{s})\in\va\De\}\in\va\TO}$ has variables
$\va\De$ of type $P\cal R$ and $\va{\TO}$ of type $PP\Si$.
Therefore, in a topos representation  it is represented by an
arrow $\Val{\{\va{s}\mid A(\va{s})\in
\va\De\}\in\va{\TO}}_\phi:P{\cal R}_\phi\times
P(P\Si_\phi)\map\O_{\tau_\phi}$. In detail (see \cite{Bell88}) we
have that
\begin{equation}
\Val{\{\va{s}\mid A(\va{s})\in \va\De\}\in\va{\TO}}_\phi =
        e_{P\Si_\phi}\circ\la \Val{\{\va{s}\mid A(\va{s})\in
\va\De\}}_\phi,\Val{\va{\TO}}_\phi\ra
\end{equation}
where $e_{P\Si_\phi}:P\Si_\phi\times
P(P\Si_\phi)\map\O_{\tau_\phi}$ is the usual evaluation arrow. In
using this expression we need the $\phi$-representatives:
\begin{eqnarray}
    \Val{\{\va{s}\mid A(\va{s})\in\va\De\}}_\phi
        :P\R_\phi&\map& P\Si_\phi\\[3pt]
    \Val{\va{\TO}}_\phi:P(P\Si_\phi) &
    \overset{\id}{\longrightarrow}& P(P\Si_\phi)
\end{eqnarray}

Finally, let $\la\name{\Xi},\name{\TO}\ra$ be a pair of global
elements in $P\R_\phi$ and $P(P\Si_\phi)$ respectively, so that
$\name{\Xi}:1_{\tau_\phi}\map P\R_\phi$ and
$\name{\TO}:1_{\tau_\phi}\map P(P\Si_\phi)$. Thus, $\name\TO$ is a
concrete truth object in $\tau_\phi$. Then, for the physical
proposition $\SAin\Xi$, we have the truth value
\begin{equation}
\TVal{\Ain\Xi}{\TO}=\Val{\{\va{s}\mid A(\va{s})\in
\va\De\}\in\va{\TO}}_\phi\la\name{\Xi},\name{\TO}\ra
:1_{\tau_\phi}\map\O_{\tau_\phi}\label{AinXiTO}
\end{equation}

\paragraph{A small generalisation.}
More generally, if $\va{K}$ and $\va\TO$ are variables of type
$P\Si$ and $P(P\Si)$ respectively, the term of interest is
`$\va{K}\in\va\TO$'. In the representation, $\phi$, of $\L{S}$,
this term maps to an arrow
$\Val{\va{K}\in\va\TO}_\phi:P\Si_\phi\times
P(P\Si_\phi)\map\O_{\tau_\phi}$.  Here,
\begin{equation}
\Val{\va{K}\in\va{\TO}}_\phi =
        e_{P\Si_\phi}\circ\la \Val{\va{K}}_\phi,
\Val{\va{\TO}}_\phi\ra
\end{equation}
where
\begin{eqnarray}
    \Val{\va{K}}_\phi
        :P\Si_\phi&\overset{\id}{\longrightarrow}& P\Si_\phi\\[3pt]
    \Val{\va{\TO}}_\phi:P(P\Si_\phi) &\overset{\id}
    {\longrightarrow}& P(P\Si_\phi)
\end{eqnarray}
Let $\name{K}$, $\name\TO$ be  global elements of $P\Si_\phi$ and
$P(P\Si_\phi)$ respectively, so that $\name{K}:1_{\tau_\phi}\map
P\Si_\phi$ and $\name\TO:1_{\tau_\phi}\map P(P\Si_\phi)$. We adopt
the notation $\la\name{K},\name\TO\ra:1_{\tau_\phi}\map
P\Si_\phi\times P(P\Si_\phi)$.  Then  the truth of the
(mathematical) proposition ``$\name{K}\in\TO$'' is
\begin{eqnarray}
\TValM{\name{K}\in\TO}&=&
        \Val{\va{K}\in\va\TO}_\phi\, \langle
        \name{K},\name\TO\rangle                \nonumber\\
        &=&e_{P\Si_\phi}\circ\la\name{K},\name{\TO}\ra:
                1_{\tau_\phi}\map \O_{\tau_\phi}
        \label{TValMgainTO}
\end{eqnarray}

\subsubsection{The Example of Classical Physics.}
If classical physics is studied this way, the general formalism
simplifies, and the term `$\{\va{s}\mid A(\va{s})\in\va{\De}\}
\in\va\TO$' is represented by the function $\Val{\{\va{s}\mid
A(\va{s})\in\va{\De}\} \in\va\TO}_\s :P\mathR\times
P(P\Si_\s)\map\O_{\Set}\simeq\{0,1\}$ defined by
\begin{eqnarray}
\TVal{\Ain\De}{\TO}= \Val{\{\va{s}\mid A(\va{s})\in\va{\De}\}
\in\va\TO}_\s(\De,\TO)
                &=&
 {\left\{\begin{array}{ll}
            1 & \mbox{\ if\ $\{s\in\Si_\s\mid A_\s(s)
                        \in\De\} \in \TO$;} \\
            0 & \mbox{\ otherwise}
         \end{array}
        \right.}\nonumber\\[5pt]
&=&
 {\left\{\begin{array}{ll}
            1 & \mbox{\ if\ $A_\s^{-1}(\De) \in \TO$;} \\
            0 & \mbox{\ otherwise}
         \end{array}
        \right.}\label{Def:nu(AinD;T}
\end{eqnarray}
for all $\TO\in P(P\Si_\s)$. We can clearly see the sense in which
the truth object $\TO$ is playing the role of a state. Note that
the result \eq{Def:nu(AinD;T} of classical physics is a special
case of \eq{AinXiTO}.

To recover the usual truth values given in  \eq{nuAinDe;s},  a
truth object, $\TO^s$, must be associated with each micro-state
$s\in\Si_\s$. The correct choice is
\begin{equation}
        \TO^s:=\{K\subseteq\Si_\s\mid s\in K\}
                        \label{Def:ClassTO}
\end{equation}
for each $s\in\Si_\s$. It is clear that $s\in A_\s^{-1}(\De)$ (or,
equivalently, $A_\s(s)\in\De$) if, and only if,
$A_\s^{-1}(\De)\in\TO^s$. Hence \eq{Def:nu(AinD;T} can be
rewritten as
\begin{equation}\label{Def:[AinD]Class(2)}
        \TVal{\Ain\De}{\TO^s}:=
        \left\{\begin{array}{ll}
            1 & \mbox{\ if\ $s\in A_\s^{-1}(\De)$;} \\
            0 & \mbox{\ otherwise.}
         \end{array}
        \right.
\end{equation}
which reproduces  \eq{nuAinDe;s} once $\TVal{\Ain\De}{s}$ is
identified with $\TVal{\Ain\De}{\TO^s}$.

\subsubsection{The Truth Object $\ps\TO^{\ket\psi}$\ in Quantum Theory}
We can now start to discuss the application of these ideas to
quantum theory.  In order to use  \eq{TValMgainTO} or \eq{AinXiTO}
we need to construct a concrete truth object, $\name{\ps\TO}$, in
the topos $\tau_\phi:=\SetH{}$. Thus
$\name{\ps\TO}:1_{\tau_\phi}\map P(P\Sig)$; equivalently, $\ps\TO$
is a sub-object of $P\Sig$.

However, we have to keep in mind the need to restrict to
\emph{clopen} sub-objects of $\Sig$. In particular, we have to
show that there is a well-defined presheaf $\PSig$ such that
\begin{equation}
        \Subcl{\Sig}\simeq\Ga(\PSig)      \label{Subcl=GPSig}
\end{equation}
We will prove this shortly in Section
\ref{SubSec:PresheafP-clSig}.

Given \eq{Subcl=GPSig},  $\ps{K}\in\Subcl\Sig$ is equivalent to an
arrow $\name{\ps{K}}:1_{\tau_\phi} \map\PSig$; and hence a truth
object, $\ps{\TO}$, has to be a sub-object of $\PSig$ in order
that the valuation $\TValM{\name{K}\in\TO}$ in \eq{TValMgainTO} is
meaningful.

This truth value, $\TValM{\name{K}\in\TO}$, is a global element of
$\Om$, and in the topos of presheaves, $\SetH{}$, we have (see
\eq{ValxinKpresheaf})
\begin{equation}
\TValM{\name{\ps{K}}\in\ps\TO}_V:=\{V^\prime\subseteq V\mid
                        \ps{K}_{V^\prime}\in\ps\TO_{V^\prime}\}
\end{equation}
for each context $V$.

There are various examples of $\ps{K}$ that are of interest to us.
In particular, let $\ps{K}=\das{P}$ for some projector $\P$. Then,
using the propositional language $\PL{S}$  discussed earlier in
this paper, the `truth' of the proposition represented by $\P$
(for example,  $\SAin\De$) is
\begin{equation}
\TValM{\name{\das{P}}\in\ps\TO}_V =\{V^\prime\subseteq V\mid
\dasto{V^\prime}{P}\big)\in\ps\TO_{V^\prime}\}\label{ValDasFinal}
\end{equation}
for all stages $V$.

In the case of the local language $\L{S}$, the important example
is when the sub-object $\ps{K}$ of $\Sig$ is of the form
$A_\phi^{-1}(\ps\Xi)$, for some sub-object $\ps\Xi$ of $\ps{R}$ .
This will yield the truth value, $\TVal{\Ain\ps\Xi}{\ps\TO}$, in
\eq{AinXiTO}. However, to discuss this further requires  the
representation of function symbols $A:\Si\map\R$ in the topos
$\SetH{}$, and this is deferred until paper III \cite{DI(3)}.

\paragraph{The definition of the truth objects
$\ps\TO^{\ket\psi}$.} The definition of truth objects in quantum
theory was studied in the original papers
\cite{IB98,IB99,IB00,IB02}. It was shown there that to each
quantum state $\ket\psi\in\cal H$, there corresponds a truth
object, $\ps\TO^{\ket\psi}$, which is defined as the following
sub-object of the outer presheaf, $\G$:
\begin{eqnarray}
\ps\TO^{\ket\psi}_V&:=&\{\hat\alpha\in \G_V\mid
                {\rm Prob}(\hat\alpha;\ket\psi)=1\}\label{TOpsi1}
                                                \nonumber\\[2pt]
        &=&\{\hat\alpha\in \G_V\mid
                \bra\psi\hat\alpha\ket\psi=1\}     \label{TOpsi2}
\end{eqnarray}
for all stages $V\in\Ob{\V{}}$. Here, ${\rm
Prob}(\hat\alpha;\ket\psi)$ is the usual expression for the
probability that the proposition represented by the projector
$\hat\alpha$ is true, given that the quantum state is the
(normalised) vector $\ket\psi$.

It is easy to see that \eq{TOpsi2} defines a genuine sub-object
$\ps\TO^{\ket\psi}=\{\ps\TO^{\ket\psi}_V\mid V\in\Ob{\V{}}\}$ of
$\G$, since if $\hat\beta\succeq\hat\alpha$, then
$\bra\psi\hat\beta\ket\psi \geq\bra\psi\hat\alpha\ket\psi$.
Therefore, if $V^\prime\subseteq V$ and $\hat\alpha\in \G_V$, then
$\bra\psi \G(i_{V^\prime V})(\hat\alpha)\ket\psi
\geq\bra\psi\hat\alpha\ket\psi$. In particular, if
$\bra\psi\hat\alpha\ket\psi=1$ then $\bra\psi \G(i_{V^\prime
V})(\hat\alpha)\ket\psi =1$.

We note that there is an interesting connection between
$\TO^{\ket\psi}$ and (outer) daseinisation. Specifically, at each
stage $V$, take the smallest projection $\hat\alpha_0(V)$ in
$\TO^{\ket\psi}_V$. Then, clearly, since
$\bra\psi\hat\alpha\ket\psi=1$ just means
$\hat\alpha\succeq\hat{P_{\ket\psi}}$, we have
\begin{equation}
         \hat\alpha_0(V)=\delta(\hat P_{\ket\psi})_V,
\end{equation}
where $\hat{P}_{\ket\psi}$ is  the projection onto the
one-dimensional subspace generated by $\ket\psi$.

The next step is to define the presheaf $\PSig$, and then show
that there is a monic arrow $\G\map \PSig$, so that $\G$ is a
sub-object of $\PSig$. Then, since $\ps\TO^{\ket\psi}$ is a
sub-object of $\G$, and $\G$ is a sub-object of $\PSig$, it
follows that $\ps\TO^{\ket\psi}$ is a sub-object of $\PSig$, as
required.

With this definition of $\ps\TO^{\ket\psi}$, the truth value,
\eq{ValDasFinal}, for the propositional language $\PL{S}$ becomes
\begin{equation}
\TValM{\name{\das{P}}\in\ps\TO^{\ket\psi}}_V =\{V^\prime\subseteq
V\mid \bra\psi\dasto{V^\prime}{P}\ket\psi=1\}
\end{equation}

It is easy to see that the definition of a truth object in
\eq{TOpsi2} can be extended to a mixed state with a density-matrix
operator $\hat\rho$:\ simply replace the definition in \eq{TOpsi2}
with
\begin{eqnarray}
\ps\TO^{\hat\rho}_V&:=&\{\hat\alpha\in \G_V\mid
                {\rm Prob}(\hat\alpha;\rho)=1\}      \nonumber\\[2pt]
        &=&\{\hat\alpha\in \G_V\mid
               {\rm tr}(\hat\rho\hat\alpha)=1\}     \label{TOrho}
\end{eqnarray}

However there is an important difference between the truth object
associated with a vector state, $\ket\psi$, and the one associated
with a density matrix, $\rho$. In the vector case, it is easy to
see that the mapping $\ket\psi\map\ps\TO^{\ket\psi}$ is one-to-one
(up to a phase factor on $\ket\psi$) so that, in principle, the
state $\ket\psi$ can be \emph{recovered} from $\ps\TO^{\ket\psi}$
(up to a phase-factor). On the other hand, there are simple
counterexamples which show  that, in general, the density matrix,
$\rho$ \emph{cannot} be recovered from $\ps\TO^{\hat\rho}$.

In a sense, this should not surprise us. The analogue of a density
matrix in classical physics is a probability measure $\mu$ defined
on the classical state space $\cal S$. Individual microstates
$s\in\cal S$ are in one-to-one correspondence with probability
measures of the form $\mu_s$ defined by $\mu_s(K)=1$ if $s\in K$,
$\mu_s(K)=0$ if $s\not\in K$.

However, one of the main claims of our programme is that any
theory `looks like' classical physics in the appropriate topos.
This suggests that, in the topos version of quantum theory, a
density matrix should be represented by some sort of measure on
the state object $\Sig$ in the topos $\tau_\phi$; and  this should
relate in some way to an `integral' of `vector truth objects'. The
development of this idea is one of the many interesting tasks for
the future.

\subsubsection{Time-dependence and the Truth Object.}
As emphasised at the end of the first paper in this series
\cite{DI(1)}, the question of time dependence depends on the
theory-type  being considered. The structure of the language
$\L{S}$ that has been used so far is such that the time variable
lies outside the language. In this situation, the time dependence
of the system can be  implemented in several  ways.

For example, we can make the truth object time dependent, giving a
family of truth objects, $t\mapsto\ps\TO^t$, $t\in\mathR$. In the
case of classical physics, with the truth objects $\ps\TO^s$,
$s\in\Si_\s$, the time evolution  comes from the time dependence,
$t\mapsto s_t$, of the microstate in accordance with the classical
equations of motion. This gives the family $t\mapsto \ps\TO^{s_t}$
of truth objects.

Something similar  happens in quantum theory, with a family
$t\mapsto \ps\TO^{\ket\psi_t}$ of truth objects, where the states
$\ket\psi_t$ satisfy the time-dependent Schr\"odinger equation.
Thus both classical and quantum truth objects belong to a
`Sch\"odinger picture' of time evolution.

It is also possible to construct a `Heisenberg picture' where the
truth object is constant but the physical quantities and
associated propositions are time dependent. We will return to this
in the next paper  when we discuss the use of unitary operators.

\subsection{The Presheaf $\PSig$}\label{SubSec:PresheafP-clSig}
\subsubsection{The Definition of $\PSig$.}
We must now show that there really is a presheaf $\PSig$.

The easiest way of defining $\PSig$ is to start with the concrete
expression for the normal power object $P\Sig$ \cite{Gol84}.
First, if $\ps{F}$ is any presheaf over $\V{}$,  define the
\emph{restriction} of $\ps{F}$ to $V$ to be the functor
$\ps{F}\!\downarrow\!V$ from the category\footnote{The notation
$\downarrow\!\!V$ means the partially-ordered set of all
subalgebras $V^\prime\subseteq V$.} $\downarrow\!\!V$ to $\Set$
that assigns to each $V_{1}\subseteq V$, the set $\ps{F}_{V_{1}}$,
and with the obvious induced presheaf maps.

Then, at each stage $V$,  $P\Sig_V$ is the set of natural
transformations from $\Sig\!\downarrow\!V$ to $\Om\!\downarrow\!
V$. These are in one-to-one correspondence with families of maps
 $\s:=\{\s_{V_{1}}:\Sig_{V_{1}}
\map\Om_{V_{1}}\mid V_{1}\subseteq V\}$, with the following
commutative diagram for all $V_2\subseteq V_1\subseteq V$:
\begin{center}
\setsqparms[1`1`1`1;1000`700]
\square[\Sig_{V_1}`\Om_{V_1}`\Sig_{V_2}`\Om_{V_2};
\s_{V_1}`\Sig(i_{V_1V_2})`\Om(i_{V_1V_2})`\s_{V_2}]
\end{center}
The presheaf maps are defined by
\begin{eqnarray}
        P\Sig(i_{V_1 V}):P\Sig_{V}&\map& P\Sig_{V_1}\\
         \s\ \ &\mapsto& \{\s_{V_2}\mid V_2\subseteq V_1\}
\end{eqnarray}
and the evaluation arrow ${\rm ev}:P\Sig\times\Sig\map\Om$, has
the form, at each stage $V$:
\begin{eqnarray}
        {\rm ev}_V:P\Sig_V\times\Sig_V&\map&\Om_V  \\
                        (\s,\l) &\mapsto& \s_V(\l)
\end{eqnarray}

Moreover, in general, given a map $\chi:\Sig_V\map\Om_V$, the
subset of $\Sig_V$ associated with the corresponding sub-object is
$\chi^{-1}(1)$, where $1$ is the unit (`truth') in the Heyting
algebra $\Om_V$.

This suggests strongly that an object, $\PSig$, in $\SetH{}$ can
be defined using the same definition of $P\Sig$ as above, except
that the family of maps  $\sigma:=\{\sigma_{V_{1}}:\Sig_{V_{1}}
\map\Om_{V_{1}}\mid V_{1}\subseteq V\}$ must be such that, for all
$V_1\subseteq V$, $\sigma_{V_1}^{-1}(1)$ is a \emph{clopen} subset
of the (extremely disconnected) Hausdorff space $\Sig_{V_1}$. It
is straightforward to check that such a restriction is consistent,
and that $\Subcl\Sig\simeq\Ga(\PSig)$ as required.

\subsubsection{The Monic Arrow From $\G$ to $\PSig$.}
\label{SubSec:MonicGPSig} We define $\iota:\G\times\Sig\map\Om$,
with the power transpose $\name{\iota}:\G\map\PSig$, as follows.
First recall that in any topos, $\tau$  there is a bijection ${\rm
Hom}_\tau(A,C^B)\simeq {\rm Hom}_\tau(A\times B,C)$, and hence, in
particular, (using $P\Sig=\Om^{\Sig}$)
\begin{equation}
        {\rm Hom}_{\SetH{}}(\G, P\Sig)\simeq
        {\rm Hom}_{\SetH{}}(\G\times\Sig,\Om).
\end{equation}

Now let $\hat\alpha\in\PV$, and let $S_{\hat\alpha}:=
\{\l\in\Sig_V \mid\l(\hat\alpha)=1\}$ be the clopen subset of
$\Sig_V$ that corresponds to the projector $\hat\alpha$ via the
spectral theorem; see \eq{Def:Salphahat}. Then we define
$\iota:\G\times\Sig\map\Om$ at stage $V$ by
\begin{equation}
 \iota_V(\hat\alpha,\l):=\{V^\prime\subseteq V\mid
    \Sig(i_{V^\prime\,V})(\l)\in
 S_{\G(i_{V^\prime\, V})(\hat\alpha)}\}\label{Def:iV}
\end{equation}
for all $(\hat\alpha,\l)\in \G_V\times\Sig_V$.

On the other hand, the basic result relating coarse-graining to
subsets of $\Sig$ is
\begin{equation}
S_{\G(i_{V^{\prime}\,V)}(\dasto{V}{\alpha})}=
        \Sig (i_{V^{\prime}\,V})(S_{\dasto{V}{\alpha}})
\end{equation}
for all $V^\prime\subseteq V$ and for all $\hat\alpha\in\G_V$. It
follows that
\begin{equation}
        \iota_V(\hat\alpha,\l):=\{V^\prime\subseteq V\mid
        \Sig(i_{V^\prime\,V})(\l)\in
        \Sig(i_{V^\prime\, V})(S_{\hat\alpha})\}
\end{equation}
for all $(\hat\alpha,\l)\in \G_V\times\Sig_V$. In this form is is
clear that $\iota_V(\hat\alpha,\l)$ is indeed a \emph{sieve} on
$V$; \ie\ an element of $\Om_V$.

The next step is to show that the collection of maps
$\iota_V:\G_V\times\Sig_V\map \Om_V$ defined in \eq{Def:iV}
constitutes a natural transformation from the object
$\G\times\Sig$ to the object $\Om$ in the topos $\SetH{}$. This
involves chasing around  a few commutative squares, and we will
spare the reader the ordeal. There is some subtlety,  since we
really want to deal with ${\rm Hom}_{\SetH{}}(\G,\PSig)$, not
${\rm Hom}_{\SetH{}}(\G,P\Sig)$; but all works in the end.

To prove that $\name{\iota}:\G\map\PSig$ is monic, it suffices to
show that the map $\name{\iota}_V:\G_V\map\PSig_V$ is injective at
all stages $V$. This is a straightforward exercise and the details
will not be given here.

The conclusion of this exercise is that, since
$\name{\iota}:\G\map\PSig$ is monic, the truth sub-objects
$\ps\TO^{\ket\psi}$ of $\G$ can also be regarded as sub-objects of
$\PSig$, and hence the truth value assignment in \eq{ValDasFinal}
is well-defined.

Finally then, for any given quantum state $\ket\psi$ the basic
proposition $\SAin\De$ can be assigned a generalised truth value
$\TVal{\Ain\De}{\ket\psi}$ in $\Ga\Om$, where $\tau:=\SetH{}$ is
the topos of presheaves over $\V{}$. This is defined at each
stage/context $V$ as
\begin{eqnarray}
\TVal{\Ain\De}{\ket\psi}_V&:=&\TValM{\name{\delta(\hat
E[A\in\De])}\in\ps\TO^{\ket\psi}}_V
                                \nonumber       \\[4pt]
                &=&  \{V^\prime\subseteq V\mid
\name{\delta\big(\hat
E[A\in\De]\big)}_{V^\prime}\in\ps\TO^{\ket\psi}_{V^\prime}\}.
\end{eqnarray}

\section{Conclusion}
In this, the second in our series of papers on topos theory and
physics, we have started the development of a topos representation
of quantum theory. In the first half of the paper, we have shown
how propositions can be represented by clopen sub-objects of the
spectral presheaf, $\Sig$, of the quantum theory. This is
equivalent to finding a topos representation of the propositional
language, $\PL{S}$, that was discussed in paper I \cite{DI(1)}. A
key ingredient in this representation is the concept of
\emph{daseinisation}, due to de Groote. By using this operation, a
projection operator $\hat E[A\in\De]$ can be mapped to a global
element $\delta(\hat E[A\in\De])$ of the outer presheaf $\G$, and,
with the aid of the monic $\G\map \PSig$, thereby to a (clopen)
sub-object of the state object $\Sig$.

As was emphasised in the Introduction, this piece of work is
something of a sideline in regard to the main  programme, which is
to find representations of the local language $\L{S}$. However, it
is a satisfying (for us) completion of the earlier work on quantum
theory and topoi.

But it is also  a useful example with which to discuss the general
problem of what lies `inside' a language, and what lies `outside'.
In the case of the language $\PL{S}$, it is clear that most
`symbols' in the theory lie outside. This includes (i) the topos;
(ii) the physical quantities, $A$, and subsets,
$\De\subseteq\mathR$, in the primitive propositions $\SAin\De$;
and (iii) the state object whose sub-objects provide the Heyting
algebra in which $\PL{S}$ is represented.

On the other hand, in a representation of the language $\L{S}$,
the only entity that necessarily lies outside the scope of the
language, is the topos in which it is represented. As we will see
in the next paper, III, the most general propositions in the
$\L{S}$-theory are represented by the sub-objects
$A_\phi^{-1}(\Xi)$, where $\Xi$ is any sub-object of the
quantity-value object $\R_\phi$ \cite{DI(3)}. In this form, all
the important physical ingredients in the theory have linguistic
precursors in the language $\L{S}$ (with the exception of the
topos $\tau_\phi$). It is this `internal' language that would be
used to manipulate  propositions `about the world' in a theory of
this type.

The second part of the present paper is concerned with the idea of
`truth objects'. These play a central role in both the $\PL{S}$
and the $\L{S}$-representations, and are the closest analogue
there is to the notion of a micro-state in classical physics. It
is likely that most topos-based theories of physics will use these
objects because of the anticipated absence of microstates;
classical physics is, of course, an exception.

The discussion of truth objects in Section
\ref{SubSub:TruthObjects} was formulated in terms of the language
$\L{S}$, which gives a clear way of understanding the two types of
`state': microstates (global elements of the state object
$\Si_\phi$), and truth objects (global elements of the power
object $P(P\Si_\phi))$.  However, the actual quantum-theory truth
objects used in Section  \ref{SubSub:TruthObjects} are the ones
given in the original papers on toposifying quantum theory, and no
work has been done on them since then.

This is clearly an area in which further research is necessary. In
particular, we would like to know if there are any generic
properties of a truth object as an element in $P(P\Si_\phi)$; \ie\
is there a theory of such things, or can any element of
$P(P\Si_\phi)$ serve as one? This is related to a question that
arises naturally when considering the specific truth objects,
$\ps\TO^{\ket\psi}$, in quantum theory: namely what, if anything,
can be said about the object
$\ps\TO^{\alpha\ket\psi+\beta\ket\phi}$ in regard to the objects
$\ps\TO^{\ket\psi}$ and $\ps\TO^{\ket\phi}$? Or, to put it another
way, is there an analogue for truth objects of the superposition
of states? This is a very important subject for future research.

\vspace{1cm} \noindent {\bf Acknowledgements} This research was
supported by grant RFP1-06-04 from The Foundational Questions
Institute (fqxi.org). AD gratefully acknowledges financial support
from the DAAD.

This work is also supported in part by the EC Marie Curie Research
and Training Network ``ENRAGE'' (European Network on Random
GEometry) MRTN-CT-2004-005616.


\begin{thebibliography}{99}

\bibitem {DI(1)} A.~D\"{o}ring and C.J. Isham.
\newblock A topos foundation for theories of physics:
    {I.} Formal languages for physics.
\newblock (2007).

\bibitem {DI(3)} A.~D\"{o}ring and C.J. Isham.
\newblock A topos foundation for theories of physics:
    {III.} Quantum theory and the representation of
    physical quantities with arrows \mbox{$\dasBo{A}:\Sig\map\SR$}.
\newblock (2007).

\bibitem {DI(4)} A.~D\"{o}ring and C.J. Isham.
\newblock A topos foundation for theories of physics:
            {IV.} Categories of Systems.
\newblock(2007).

\bibitem {Kess07} S.~Kessari.
\newblock Affine histories in quantum gravity: Introduction
and the representation for a cosmological model.
\newblock \emph{Classical and Quantum Grav.} {\bf24}, 1303-1329, (2007).


\bibitem {Bell88} J.L.~Bell.
\newblock {\em Toposes and Local Set Theories.}
\newblock Clarendon, Oxford (1988).

\bibitem {Gol84}R.~Goldblatt.
\newblock {\em Topoi: The Categorial Analysis of Logic}.
\newblock North-Holland, London (1984).

\bibitem {KR83a} R.V.~Kadison and J.R.~Ringrose.
\newblock {\em Fundamentals of the Theory of Operator Algebras,
            Volume 1: Elementary Theory}.
            \newblock Academic Press, New York (1983).

\bibitem {Doe05} A.~D\"{o}ring.
\newblock Kochen-Specker theorem for von Neumann algebras.
\newblock {\em Int.\ J.\ Theor.\ Phys.} {\bf 44}, 139--160, (2005).

\bibitem {KS67} S.~Kochen and E.P.~Specker.
\newblock The problem of hidden variables in quantum mechanics.
\newblock {\em Journal of Mathematics and Mechanics} {\bf 17},
59--87, (1967).

\bibitem {deG05}H.F.~de~Groote. \newblock Observables.
\newblock arxiv.org/abs/math-ph/0507019, (2005).

\bibitem {deG05c}Hans F. de Groote. Observables I: Stone spectra.
\newblock arxiv.org/abs/math-ph/0509020, (2005).

\bibitem {deG04}H.F.~de~Groote. \newblock On a canonical lattice
structure on the effect algebra of a von Neumann algebra.
\newblock arxiv.org/abs/math-ph/0410018 v2, (2004).


\bibitem {IB98} C.J.~Isham and J.~Butterfield.
\newblock A topos perspective on the {K}ochen-{S}pecker theorem:
            {I.} {Q}uantum states as generalised
valuations. \newblock {\em Int.\ J.\ Theor.\ Phys.} \textbf{37},
2669--2733, (1998).

\bibitem {IB99} C.J.~Isham and J.~Butterfield.
\newblock A topos perspective on the {K}ochen-{S}pecker theorem:
    {II.} {C}onceptual aspects, and classical analogues.
\newblock  {\em Int.\ J.\ Theor.\ Phys.} \textbf{38}, 827--859, (1999).

\bibitem {IB00} J.~Hamilton, J.~Butterfield and C.J.~Isham.
\newblock A topos perspective on the {K}ochen-{S}pecker theorem:
    {III.} {V}on {N}eumann algebras as the base category.
\newblock  {\em Int.\ J.\ Theor.\ Phys.} \textbf{39}, 1413-1436, (2000).

\bibitem {IB02} J.~Butterfield and C.J.~Isham.
\newblock {A topos perspective on the Kochen-Specker theorem:
        {IV.} {I}nterval valuations}.
\newblock {\em Int.\ J.\ Theor.\ Phys.} {\bf 41}, 613--639, (2002).
\end{thebibliography}
\end{document}